\documentclass[fleqn,usenatbib,usedcolumn]{mnras}

\pdfoutput=1

\usepackage{newtxtext,newtxmath}
\renewcommand{\la}{\lesssim} 
\renewcommand{\ga}{\gtrsim} 

\usepackage[T1]{fontenc}

\usepackage{graphicx}	
\usepackage{amsmath}	
\usepackage[usenames,dvipsnames]{color}

\newcommand{\Sref}[1]{Section \ref{#1}}
\newcommand{\Tref}[1]{Table \ref{#1}}

\sfcode`\.=1001\sfcode`\?=1001\sfcode`\!=1001
\newcommand{\Fref}[1]{\ifhmode \ifnum\spacefactor=1001 Figure \ref{#1}\else Fig.\ \ref{#1}\fi \else Figure \ref{#1}\fi}
\newcommand{\Eref}[1]{\ifhmode \ifnum\spacefactor=1001 Equation (\ref{#1})\else equation (\ref{#1})\fi \else Equation (\ref{#1})\fi}

\newcommand{\ms}{\ensuremath{\textrm{m\,s}^{-1}}}
\newcommand{\kms}{\ensuremath{\textrm{km\,s}^{-1}}}

\newcommand{\SN}{\ensuremath{\textrm{S/N}}}

\newcommand{\zab}{\ensuremath{z_\textrm{\scriptsize abs}}}

\newcommand{\tran}[3]{\ensuremath{\ion{#1}{#2}\,\lambda\textrm{#3}}}

\newcommand{\daa}{\ensuremath{\Delta\alpha/\alpha}}
\newcommand{\msnm}{\ensuremath{\ms\,\textrm{nm}^{-1}}}

\usepackage{array}
\newcolumntype{:}{>{\global\let\currentrowstyle\relax}}
\newcolumntype{;}{>{\currentrowstyle}}

\title[Subaru limits on varying $\alpha$]{Subaru Telescope limits on cosmological variations in the fine-structure constant}

\author[M. T. Murphy, K. L. Cooksey]{Michael T. Murphy,$^{1}$\thanks{E-mail: mmurphy@swin.edu.au (MTM)} Kathy L. Cooksey,$^{2}$\\
  $^{1}$Centre for Astrophysics and Supercomputing, Swinburne University of Technology, Hawthorn, Victoria 3122, Australia\\
  $^{2}$University of Hawai`i at Hilo, 200 West K\={a}wili Street, Hilo, HI 96720, USA
}

\date{Accepted ---. Received ---; in original form ---}

\pubyear{2017}

\volume{{\rm in press}}

\begin{document}
\label{firstpage}
\pagerange{\pageref{firstpage}--\pageref{lastpage}}
\maketitle

\begin{abstract}
Previous, large samples of quasar absorption spectra have indicated some evidence for relative variations in the fine-structure constant ($\Delta\alpha/\alpha$) across the sky. However, they were likely affected by long-range distortions of the wavelength calibration, so it is important to establish a statistical sample of more reliable results, from multiple telescopes. Here we triple the sample of $\Delta\alpha/\alpha$ measurements from the Subaru Telescope which have been `supercalibrated' to correct for long-range distortions. A blinded analysis of the metallic ions in 6 intervening absorption systems in two Subaru quasar spectra provides no evidence for $\alpha$ variation, with a weighted mean of $\Delta\alpha/\alpha=3.0\pm2.8_{\rm stat}\pm2.0_{\rm sys}$ parts per million (1$\sigma$ statistical and systematic uncertainties). The main remaining systematic effects are uncertainties in the long-range distortion corrections, absorption profile models, and errors from redispersing multiple quasar exposures onto a common wavelength grid. The results also assume that terrestrial isotopic abundances prevail in the absorbers; assuming only the dominant terrestrial isotope is present significantly lowers $\Delta\alpha/\alpha$, though it is still consistent with zero. Given the location of the two quasars on the sky, our results do not support the evidence for spatial $\alpha$ variation, especially when combined with the 21 other recent measurements which were corrected for, or resistant to, long-range distortions. Our spectra and absorption profile fits are publicly available.
\end{abstract}

\begin{keywords}
line: profiles -- instrumentation: spectrographs -- quasars: absorption lines -- cosmology: miscellaneous -- cosmology: observations
\end{keywords}



\section{Introduction}\label{s:intro}

Absorption lines from metallic ions in gas clouds along the sight-lines to background quasars have been used to limit relative variations in the fine-structure constant ($\alpha\equiv e^2/\hbar c$) to a few parts-per-million (ppm) over cosmological time and distance scales \citep[e.g.][]{Webb:1999:884,Murphy:2001:1208,Murphy:2003:609,Quast:2004:L7,Levshakov:2005:827,Levshakov:2007:1077,Molaro:2008:173,Murphy:2008:1053,Webb:2011:191101,Agafonova:2011:28,King:2012:3370,Molaro:2013:A68,Songaila:2014:103}. A relative difference, $\daa\equiv(\alpha_z-\alpha_0)/\alpha_0$, between $\alpha$ in an absorber at redshift $z$ ($\alpha_z$) and the current laboratory value ($\alpha_0$), should be observed as a shift in the frequency (or wavenumber, $\omega_i$), and hence observed velocity, of a transition, $i$:
\begin{equation}\label{e:daa}
\frac{\Delta v_i}{c} \approx -2\frac{\Delta\alpha}{\alpha}\frac{q_i}{\omega_i}\,,
\end{equation}
where the $q_i$ coefficient characterises the sensitivity of transition $i$'s wavenumber to a varying $\alpha$ \citep{Dzuba:1999:888,Webb:1999:884}. Because different transitions have different sensitivities to $\alpha$-variation (i.e.\ different $q$ coefficients), one can measure \daa\ and $z$ simultaneously by analysing two or more transitions. This assumes that the redshifts of those transitions are the same (i.e.\ that they are not kinematically distinct), and that the relative wavelength calibration of the quasar spectra is accurate, to the degree of precision allowed by the signal-to-noise ratio (\SN) and spread in $q$ coefficients of the transitions available.

Using this ``many multiplet'' (MM) method, large samples of quasar spectra from the Keck Observatory and Very Large Telescope (VLT) have even shown some evidence for variations in $\alpha$ \citep[e.g.][]{Webb:1999:884,Murphy:2001:1208,Murphy:2003:609,Webb:2011:191101}. However, the results from the two telescopes disagree: at redshifts 0.4--4.2, 143 measurements with the Keck telescope indicated a smaller $\alpha$ \citep[$\daa=-5.7\pm1.1_{\rm stat}$\,ppm;][]{Murphy:2004:131}, while 153 measurements with VLT spectra indicated a (marginally) larger $\alpha$ \citep[$\daa=2.1\pm1.2_{\rm stat}$\,ppm;][]{King:2012:3370}. Intriguingly though, when these datasets from northern and southern telescopes are combined, there is internally consistent evidence, at the $\approx$4$\sigma$ level, for a spatial variation in $\alpha$ across the sky, the simplest model of which is a dipolar variation $\daa=(10.2\pm2.1)\,\cos(\Theta)$\,ppm, where $\Theta$ is the angle from the pole direction ${\rm (RA,Dec.)}=(17.4\pm0.9\,{\rm h},-58\pm9\,{\rm deg})$ \citep{Webb:2011:191101,King:2012:3370}. An important observational challenge remains, of course, to obtain new, independent results with high enough precision and accuracy to test these surprising results.

However, important systematic errors have recently been discovered in the high-resolution, slit-based spectrographs on the Keck, VLT and also Subaru telescopes, manifesting as long-range distortions in the wavelength scales of quasar spectra \citep[e.g.][]{Rahmani:2013:861,Bagdonaite:2014:10,Evans:2014:128,Songaila:2014:103,Whitmore:2015:446,Kotus:2017:3679}. This is despite earlier studies finding no evidence for distortions of a similar size \citep{Molaro:2008:559,Griest:2010:158,Whitmore:2010:89}, suggesting that the distortions vary over time (or other observational parameters). Indeed, throughout $\sim$15 years of Keck and VLT archival spectra, \citet{Whitmore:2015:446} found that the distortions varied substantially, with linear slopes ranging between $-3$ and $8$\,\msnm. Assuming a very simple model for these distortions and, importantly, the same slope for all exposures, \citet{Whitmore:2015:446} found the distortions to be a plausible explanation of the \citet{King:2012:3370} VLT results. However, a similar constant-slope model could only partially explain the \citet{Murphy:2003:609,Murphy:2004:131} Keck results. \citet{Dumont:2017:1568} recently found that a more complex, more realistic distortion model may be required for VLT spectra and may even allow the distortion slope and \daa\ to be determined simultaneously in some cases, but a constant slope for all exposures is still assumed. Nevertheless, by comparing multiple exposures (or combined epochs of exposures) of individual quasars, \citet{Evans:2014:128} and \citet{Kotus:2017:3679} confirmed large relative distortions between exposures that, left uncorrected, would have caused considerable spurious systematic errors in \daa\ of up to 10--15\,ppm.

While the precise effect of the long-range distortions on the large Keck and VLT samples -- and, therefore, any dipolar spatial variation in $\alpha$ -- remains unclear, it is important to establish new, reliable measurements of \daa\ in a sample of spectra or absorbers that are corrected for, or insensitive to, this effect. \citet{Evans:2014:128}, \citet{Murphy:2016:2461} and \citet{Kotus:2017:3679} reported the first results of this effort: 21 separate measurements in 13 absorbers using Keck, Subaru and VLT which do not indicate any variations in $\alpha$ at the 1\,ppm level (the weighted mean result is $\daa = -1.2\pm0.5_{\rm stat}\pm0.5_{\rm sys}$). However, the 11 quasars studied are all relatively close on the sky to the equator of the \citet{King:2012:3370} dipole model, i.e.\ \daa\ is not expected to deviate substantially from zero for these quasars in that model. Therefore, despite the possibility that the dipolar model may be undermined by long-range wavelength distortions, an important and completely independent test is possible if new measurements of \daa\ focus, as much as possible, on quasars closer to the pole and anti-pole of that model where larger \daa\ deviations are expected.

Finally, we note that only one Subaru spectrum has been utilised for $\alpha$ variation studies so far \citep{Evans:2014:128}. It is important to study spectra from all available telescopes to discover any further systematic errors and, assuming any remaining, undiscovered effects differ from telescope to telescope, to help reduce their impact on average \daa\ results.

With these motivations, two quasars were selected for this work, PG\,0117$+$213 (hereafter J0120$+$2133) and HS\,1946$+$7658 (hereafter J1944$+$7705). These are relatively bright \citep[UK $R$ magnitudes in the SuperCosmos survey of $\approx$15.6\,mag;][]{Hambly:2001:1279}, so a high \SN\ can be built up in relatively short observations to constrain \daa\ with $\sim$3--10\,ppm precision. This is competitive with other recent measurements that were corrected for, or insensitive to, long-range distortions. These quasars also lie closer than other recent measurements to the anti-pole (within 57 degrees) of the dipole model, making them more important for an independent check of that possibility. \Sref{s:obs} details our observations of these two quasars, the data reduction and, most importantly, ``supercalibration'' procedure to correct the long-range distortions. \Sref{s:analysis} describes our analysis of 7 absorption systems in these quasar spectra, 6 of which provide new \daa\ measurements. The results are presented in \Sref{s:res}, including estimates of the most important systematic uncertainties. These results are compared with other recent measurements in \Sref{s:disc} and combined to test the dipole model. \Sref{s:conclusion} provides our conclusions.

\section{Observations, data reduction and calibration}\label{s:obs}

\subsection{Quasar and solar twin observations}\label{ss:obs}

J0120$+$2133 was observed with the Subaru Telescope's High Dispersion Spectrograph \citep[HDS;][]{Noguchi:2002:855} in two runs -- 2012 December (1 night) and 2014 August (3 nights) -- in seven $\approx$3300-s exposures for a total of $\approx$6.5\,h, and J1944$+$7705 was observed only during the 2014 run in three 3300-s exposures (total $\approx$2.8\,h). Details of the two runs are summarised in \Tref{t:obs}. Observing conditions varied considerably during the 2012 run, with increasing cloud cover preventing further observations, and variable, albeit relatively low, seeing of 0.5--1.0\,arcsec. By contrast, the 2014 run was clear with relatively stable full-width-at-half-maximum (FWHM) seeing of 0.6--0.8\,arcsec during the first 2 nights but highly variable, poor seeing (0.8--2.5\,arcsec) during the final night.

\begin{table*}
\begin{center}
  \caption{Quasar observation details. The quasar name, the J2000 right-ascension and declination are provided in the first three columns and the fourth column provides the universal date and time of the start of each exposure. The fifth column is the exposure time. The two standard wavelength settings (sixth column), Yb and Yd, were employed, and these nominally cover 484--681 and 408--673\,nm (with inter-chip gaps at 545--563 and 535--553\,nm), respectively. The slit width projected on the sky (seventh column) was kept constant throughout both runs and was normally comparable to (though normally marginally smaller than) the FWHM seeing (eighth column). The \SN\ of each extracted exposure at 4300, 5000 and 6000\,\AA\ is provided in the final three columns. When combined (see \Sref{ss:red}), the \SN\ values, per 1.75-\kms\ pixel, for the master spectra at these three wavelengths are 53, 76 and 63 for J0120$+$2133, and 31, 53 and 52 for J1944$+$7705, respectively.}
\label{t:obs}
{\footnotesize
\begin{tabular}{lcccccccccc}\hline
Object       & RA [hr]    & Dec.\ [deg]    & Universal time        & Exposure & Setting & Slit     & Seeing   & \multicolumn{3}{c}{\SN\,pix$^{-1}$ at $\lambda_{\rm obs}\sim$} \\
             & \multicolumn{2}{c}{(J2000)} & [yyyy-mm-dd hh:mm:ss] & [s]      &         & [arcsec] & [arcsec] & 4300\,\AA & 5000\,\AA & 6000\,\AA   \\\hline
J0120$+$2133 & 01:20:17.3 & $+$21:33:46    & 2012-12-02 04:42:24   & 3300     & Yd      & 0.8      & 0.8--1.0 & 14        & 27        & 31          \\
             &            &                & 2012-12-02 05:39:28   & 3447     & Yd      & 0.8      & 0.6--0.8 & 18        & 27        & 26          \\
             &            &                & 2012-12-02 07:29:42   & 3300     & Yd      & 0.8      & 0.6--0.8 & 16        & 22        & 25          \\
             &            &                & 2014-08-13 10:59:10   & 3300     & Yd      & 0.8      & 0.8      & 27        & 35        & 27          \\
             &            &                & 2014-08-14 12:36:50   & 3300     & Yd      & 0.8      & 0.8      & 23        & 30        & 26          \\
             &            &                & 2014-08-14 13:41:55   & 3300     & Yd      & 0.8      & 0.8      & 22        & 29        & 36          \\
             &            &                & 2014-08-15 13:32:27   & 3300     & Yd      & 0.8      & 0.9--1.1 & 22        & 29        & 25          \\
J1944$+$7705 & 19:44:55.0 & $+$77:05:52    & 2014-08-13 08:35:32   & 3300     & Yb      & 0.8      & 0.8--1.0 & 20        & 33        & 32          \\
             &            &                & 2014-08-13 09:48:08   & 3300     & Yb      & 0.8      & 0.8--1.1 & 20        & 34        & 34          \\
             &            &                & 2014-08-15 07:30:35   & 3300     & Yb      & 0.8      & 1.2--2.5 & 14        & 24        & 25          \\
\hline
\end{tabular}
}
\begin{minipage}{\textwidth}
\end{minipage}
\end{center}
\end{table*}

The most important aspect of the observations for measuring \daa\ is the wavelength calibration. An initial calibration of each quasar exposure was established from an ``attached'' thorium--argon (ThAr) comparison lamp exposure -- i.e.\ immediately following the quasar exposure without any changes to the spectrograph set-up or telescope position. To ``supercalibrate'' this initial calibration -- i.e.\ correct it for long-range distortions -- a solar twin star was observed, with its own attached ThAr exposure, immediately following the quasar$+$ThAr pair of exposures in the 2014 run, again without any intervening change to the spectrograph set-up. However, for the quasar exposures in the 2012 run (J0120$+$2133 only) the supercalibrations were not ``attached''; instead, supercalibration exposures of two asteroids (Ceres and Vesta), taken up to 6 hours after the first 2012 quasar exposure, are used below (\Sref{ss:scal}) to understand the general size and variability of the long-range distortions in that run.

For J0120$+$2133 the very bright ($V\approx7.4$\,mag), very nearby (6$^\circ$ away) solar twin HD005294 was selected for supercalibration. For J1944$+$7705, the nearest suitable solar twin (to our knowledge) is HD186408, which is over 26$^\circ$ away. However, because the ultimate cause of the long-range distortions is not known, we planned to test the possibility that they depend strongly on telescope position by observing both HD1866408 and another solar twin, HD197027 ($V=9.2$\,mag), which is 105$^\circ$ away from J1944$+$7705 (and also well separated from the other solar twins). Unfortunately, only exposures of HD197027 were obtained. Nevertheless, together with the observations of J0120$+$2133's solar twin (HD005294), we find that the long-range distortions do not seem to depend strongly on telescope position, so the HD197027 exposures were used to supercalibrate the J1944$+$7705 observations. The solar twin exposures were very short (7--20\,s) to avoid saturating the charge coupled device (CCD) but still produce an extracted spectrum with $\SN\ga100$\,per 1.75-\kms\ pixel at 5500\,\AA.

\subsection{Data reduction, artefact removal and exposure combination}\label{ss:red}

There is currently no dedicated data reduction pipeline for Subaru/HDS spectra, so we reduced all spectra with the general suite of echelle reduction tools within {\sc iraf}\footnote{{\sc iraf} is distributed by the National Optical Astronomy Observatory, which is operated by the Association of Universities for Research in Astronomy (AURA) under cooperative agreement with the National Science Foundation.} in a similar manner to that described in \citet{Evans:2014:128}. The standard processes were used for overscan subtraction and corrections for CCD non-linearity, flat field and scattered light, and we accurately traced the quasar flux in all echelle orders using a bright standard star exposure. The blaze function was estimated and removed from the quasar spectra using a combined flat-field image extracted using the quasar's trace information. Most care was paid to the wavelength calibration steps. The ThAr flux was extracted with the corresponding quasar's trace information and the wavelength solution was established with a custom 2-dimensional polynomial fit using the \citet{Murphy:2007:221} ThAr line-list as input.

For each quasar, the extracted flux from all exposures was combined with variance weighting using {\sc uves\_popler} \citep[][version 0.73]{Murphy:2016:UVESpopler} to form a final, 1-dimensional ``master spectrum'', including a 1$\sigma$ uncertainty spectrum. \citet{Evans:2014:128} and \citet{Murphy:2016:2461} describe the details of this process. Briefly, for our HDS spectra, the most important aspects are: (i) each echelle order, of each exposure, was redispersed onto a common, log-linear vacuum--heliocentric wavelength grid with a 1.75\,\kms\,pix$^{-1}$ dispersion; (ii) after initial, automatic `cleaning' of the spectra, artefacts in the flux arrays, such as `cosmic rays', `ghosts' (internal reflections within the spectrograph), spurious or outlying pixels, and residual blaze removal effects, were masked and/or normalised out manually; these steps were recorded and are fully reproducible; and (iii) an initial, automatically set, polynomial continuum is manually refitted, if required, in regions where metal absorption is present, to ensure profile fits can be established as accurately as possible.

The resulting \SN\ of the J0120$+$2133 master spectrum is 53, 76, 63\,pix$^{-1}$ at 4300, 5000 and 6000\,\AA\ respectively. For J1944$+$7705, the \SN\ is 31, 53 and 52\,pix$^{-1}$ at the same wavelengths. The nominal resolving power of HDS with a uniformly-illuminated 0\farcs8-wide slit is $\approx$45,000. However, for the majority of our exposures, the FWHM seeing was similar to or smaller than this slit width (\Tref{t:obs}), so the real resolving power will be higher. To correct for this we followed the approach of, e.g.\ \citet{Kotus:2017:3679}: one-dimensional models of how the slit truncates Gaussian seeing profiles indicate that the real resolving power is up to 20--25\% larger than the nominal value when the seeing FWHM and slit width are approximately equal. Therefore, we used a larger resolving power of 55,000 when fitting the absorption lines in \Sref{s:analysis}. The extracted flux and uncertainty spectra from each exposure, the master spectra and the {\sc uves\_popler} log files (used to create the latter from the former), are all publicly available in \citet{Murphy:2017:alphaSubaru}.

\subsection{Supercalibration and slit-shift correction}\label{ss:scal}

\begin{figure}
\begin{center}
\includegraphics[width=1.0\columnwidth]{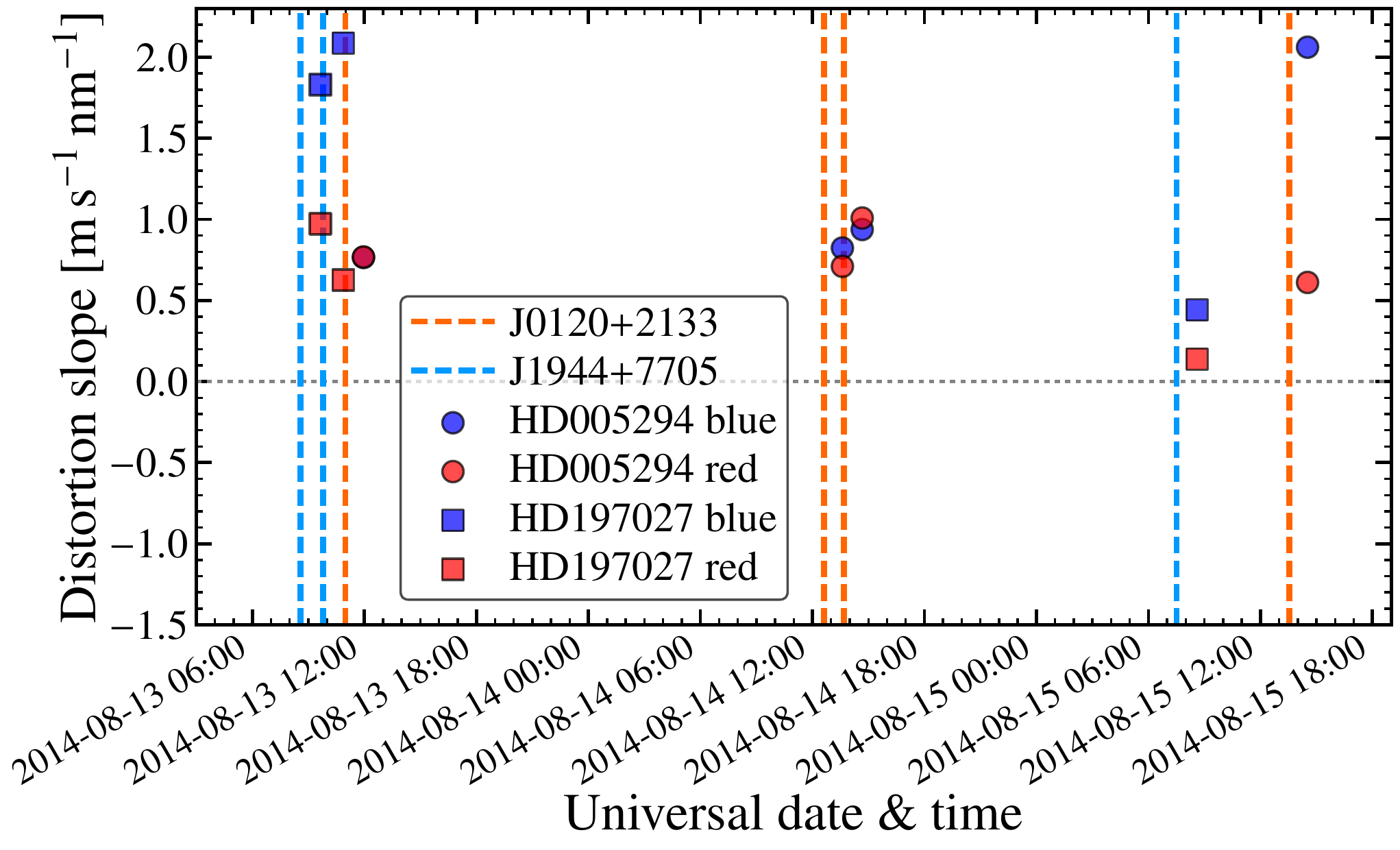}\vspace{-1em}
\caption{Supercalibration results from the 2014 run. Each $\approx$50-min quasar exposure's starting time is shown by a vertical dashed line, and is immediately followed by the corresponding solar twin supercalibration slope measurement (circles for J0120$+$2133 and HD005294, squares for J1944$+$7705 and HD197027; blue shading for the blue chip and red shading for the red chip). Note that the supercalibration slopes for the blue and red chips of the first J0120$+$2133 exposure are very similar; they are not distinguishable in the plot. The uncertainty for individual slope measurements is $\sim$0.3\,\msnm\ but the slope can vary by $\la$0.5\,\msnm\ over one-hour timescales and this sets the systematic uncertainty for the distortion correction to the quasar exposures (see \Sref{sss:long}).}
\label{f:scal_Aug14}
\end{center}
\end{figure}

The solar twin supercalibration exposures were processed with the same method detailed by \citet{Whitmore:2015:446}. Briefly, the solar twin exposures were reduced and wavelength calibrated in the same way as the quasar exposures (\Sref{ss:red} above) and each 500\,\kms-wide section of each echelle order's extracted spectrum was then compared with a solar spectrum model derived from a more accurately calibrated Fourier transform spectrum (FTS) of the Sun \citep[``KPNO2010'' from][]{Chance:2010:1289}. This comparison yields a best-fit velocity shift between the observed and FTS spectra for each chunk. As in previous works using this technique \citep[e.g.][]{Whitmore:2015:446,Dapra:2016:192,Dapra:2017:3848}, these velocity shifts showed an approximately linear long-range trend with wavelength together with much shorter-range distortions in a pattern that repeated from order to order within the same exposure. We refer to the former as ``long-range distortions'' and the latter as ``intra-order distortions''. A least-squares fit of the velocity shift information as a function of wavelength yields the long-range distortion slope for each chip (blue and red) of each exposure. These distortion slopes are plotted in \Fref{f:scal_Aug14} for the 2014 observing run.

\begin{figure}
\begin{center}
\includegraphics[width=1.0\columnwidth]{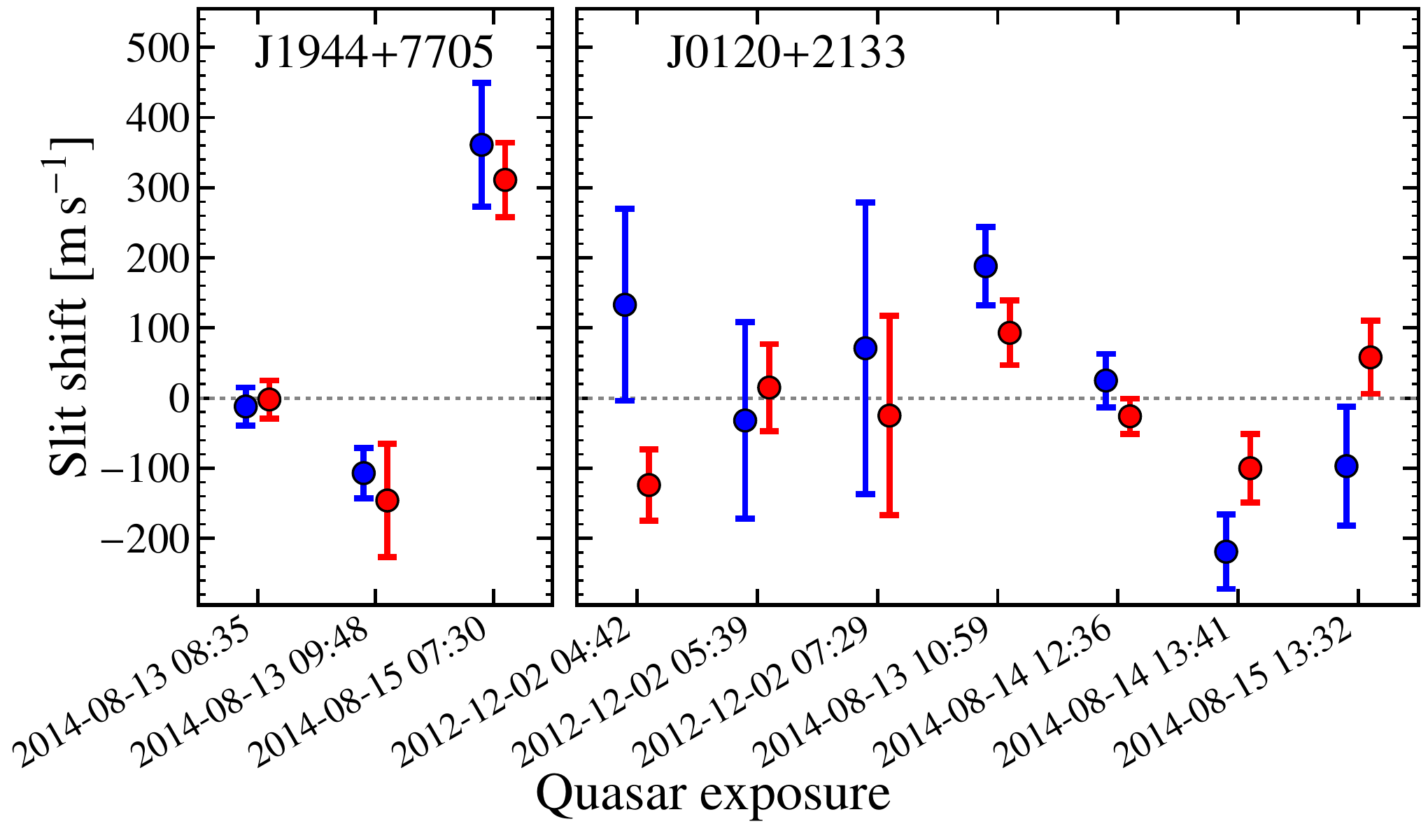}\vspace{-1em}
\caption{Slit shifts for all quasar exposures. The slit shifts for the blue and red chips (blue and red circles, respectively) are strongly correlated, and vary by the equivalent of $\la$10\% of the slit width ($\sim$400\,\ms), as expected for shifts generated by differences in alignment of the quasar within the spectrograph slit for different exposures.}
\label{f:slit_shifts}
\end{center}
\end{figure}

For J1944$+$7705, a solar twin supercalibration exposure was taken immediately after each quasar exposure. Therefore, the distortion slopes derived from the supercalibration exposures in \Fref{f:scal_Aug14} are assumed to apply directly to the corresponding quasar exposures. We assess the likely systematic errors involved in \Sref{sss:long}. The quasar exposures are then re-combined in the same way as before using {\sc uves\_popler} but now with their wavelength scales corrected for the linear distortion slopes. We then use this distortion-corrected spectrum to determine and correct for ``slit shifts'': velocity shifts between exposures caused primarily by differences in alignment of the quasar within the spectrograph slit during the observations. The slit shifts are determined using the direct comparison (DC) method introduced by \citet{Evans:2013:173} applied to each exposure and the distortion-corrected combination of all exposures: the spectral features in the spectra determine a best-fit velocity shift and uncertainty between them as a function of wavelength (in wavelength ranges where they overlap). The slit shift for each chip (blue and red) is the weighted mean of these velocity shifts where the $\chi^2$ around the weighted mean is forced to unity by adding a constant error term in quadrature to all velocity shift uncertainties; this term is generally small and is consistent with the size of the intra-order distortions observed in the supercalibration spectra (i.e.\ $\sim\pm$100\,\ms, see \Sref{sss:intra}), as expected. \Fref{f:slit_shifts} shows the results for each chip in each exposure. The slit shifts for the blue and red chip of the same quasar exposure are strongly correlated, as expected, and we observe a variation between exposures of $\sim$400\,\ms, which is expected from differences in slit alignment of $\la$10\% of the 0\farcs8 slit width.

The DC method comparison also serves as a consistency check on the long-range distortion corrections. We did not observe any significant trend in the DC velocity shifts between exposures as a function of wavelength. However, because the \SN\ of individual quasar exposures is only $\sim$30\,pix$^{-1}$ (\Tref{t:obs}), the uncertainty on this DC slope is relatively large: 0.85--3.2\,\msnm\ for both the blue and red chips. Nevertheless, this is a confirmation that the long-range distortions do not depend strongly on the telescope position, as HD197027 is over 100\,degrees away from J1944$+$7705. After the slit shifts are determined, the exposures are re-combined again, this time with both the long-range distortions and slit-shifts corrected for, and this forms the spectrum we use for subsequent spectral analysis. We also conduct the DC method analysis again with this final spectrum as a consistency check; as expected, we do not find significant slopes or offsets between the corrected exposures and the combined, corrected spectrum.

For J0120$+$2133, only the 4 exposures from the 2014 run have attached solar twin exposures, while in 2012 asteroid supercalibration exposures were taken up to 6 hours after the first quasar exposure. Our general approach is therefore to first correct the 2014 exposures for long-range distortions and slit shifts, and then treat the 2012 exposures. Specifically, we performed the same analysis for the 2014 exposures as performed above for those of J1944$+$7705 except that, instead of using the DC method to compare individual exposures with the combined spectrum from all exposures, the comparison was instead with a ``sub-spectrum'' formed only from the 2014 exposures: after all exposures were combined in {\sc uves\_popler}, including automatic and manual artefact removal and continuum fitting, the 2012 exposures were removed and the 2014 exposures were combined again. The resulting long-range distortions and slit shifts are plotted in Figs.\ \ref{f:scal_Aug14} and \ref{f:slit_shifts}. As with J1944$+$7705, we find that the slit shifts in the blue and red chips are strongly correlated and vary by $\sim$300\,\ms, as expected.

For the 3 exposures of J0120$+$2133 in 2012, we first used the DC method to compare their sub-spectrum to the corrected 2014 sub-spectrum. No significant distortion between these sub-spectra was observed. However, the uncertainty in the distortion slope was 4.8 and 2.9\,\msnm\ for the blue and red chips, respectively. The non-attached asteroid supercalibration exposures, taken up to 6 hours after the quasar exposures, have distortions slopes of 100--190\,\ms\ in the red chip, where most transitions used in this analysis fall, and 190--290\,\ms\ for the blue chip. Therefore, our approach is not to make a distortion slope correction for the 2012 quasar exposures; while we do not expect very large distortions ($\ga$2.9\,\msnm), we must account for a possible distortion slope of $\sim$2.0\,\msnm\ when determining the systematic uncertainty budgets for our \daa\ measurements from the J0120$+$2133 spectrum in \Sref{sss:long}.

The slit shifts for the 2012 exposures were determined in a similar way to the 2014 J0120$+$2133 exposures except that, first, an offset between the 2012 and 2014 sub-spectra of 500\,\ms\ was removed. This was determined in the initial DC method comparison referred to above. The slit shifts plotted in \Fref{f:slit_shifts} for the 2012 exposures are relative to the 2012 sub-spectrum and do not include this 500\,\ms\ offset. Once corrected for these shifts, all exposures of J0120$+$2133 were re-combined to form the spectrum used in subsequent spectral analysis. As with J1944$+$7705, we performed the DC method comparisons again, relative to this corrected spectrum (and its 2012 and 2014 sub-spectra), to check that no significant distortions or shifts remained.

\section{Analysis}\label{s:analysis}

\subsection{Profile fitting and \boldmath{$\daa$} measurement approach}\label{ss:genfit}

We follow a very similar approach to fitting profiles to the absorption lines and deriving \daa\ as several previous MM analyses \citep[e.g.][]{Murphy:2001:1223,Murphy:2003:609,King:2012:3370,Molaro:2013:A68}. In particular, our approach is the same as in \citet{Evans:2014:128} and \citet{Murphy:2016:2461}, so we refer the reader to these works for detailed discussion and provide only a brief summary here.

We use non-linear least-squares $\chi^2$ minimisation software, {\sc vpfit} \citep[version 9.5;][]{Carswell:2014:VPFIT}, to fit multiple velocity components to all transitions considered in an absorption system. Many models with (generally) an increasing number of components are trialled with \daa\ fixed to zero. The three parameters characterising each Voigt profile component -- the redshift ($z$), Doppler $b$-parameter and column density ($N$) -- are identical for transitions of the same ionic species (e.g.\ \ion{Fe}{ii}). The redshifts of corresponding components in different species are tied together (i.e.\ they are the same parameter), which is required by \Eref{e:daa} in order to derive \daa. The $b$-parameters are also tied together, corresponding to the assumption of a turbulent line broadening mechanism (see discussion below), and limited to $>$1\,\kms\ to avoid components becoming much narrower than the spectral resolution ($\approx$6\,\kms) and greatly slowing down {\sc vpfit}'s $\chi^2$ minimisation convergence. We select the model with the lowest $\chi^2$ per degree of freedom, $\chi^2_\nu$, for subsequent analysis \citep{Murphy:2008:1053}.

It is important to note that, until this stage of the analysis, the quasar spectra were `blinded': small, artificial long-range and intra-order distortions were introduced with random magnitudes which were unknown to the user within {\sc uves\_popler}. This was to further avoid human bias in establishing and selecting the profile model from which \daa\ is determined (even though our experience is that this is not possible with the above modelling approach). The maximum magnitudes of the distortions are small enough not to cause noticeable (by eye) shifts between transitions while various models are being trialled, but they are large enough to cause significant, spurious shifts in \daa\ when it is derived from the combined spectrum. Different distortions are applied to the different exposures but all distortions carry a common element so that they do not average out in the combined spectrum (from which \daa\ is determined). After the best-fitting profile model is selected, the spectrum is `unblinded' in {\sc uves\_popler} and no further change to the velocity structure is made in the subsequent analysis, avoiding any possible bias. {\sc vpfit} is used to minimize $\chi^2$ between the selected profile model and the unblinded spectrum; the resulting model is referred to as the ``fiducial'' one for a particular absorber, and is the starting point for all subsequent measurements and tests for systematic errors.

\daa\ is introduced as a free parameter only after the fiducial model is determined from the unblinded spectrum. {\sc vpfit} minimizes $\chi^2$ by varying all free parameters (including $z$, $b$ and $N$ for all velocity components) until $\chi^2$ differs between iterations by less than $\approx$1$\times10^{-3}$. The statistical uncertainty on \daa\ is computed from the relevant diagonal term of the covariance matrix and is multiplied by a factor of $\sqrt{\chi^2_\nu}$ to correct for a final $\chi^2_\nu$ value larger than unity. The final fit, and corresponding \daa\ value, is only accepted if the following selection criteria are met: (i) $\chi^2_\nu<1.5$; (ii) the residuals between the spectral data and profile fit show no long-range ($\ga$5 pixel), coherent deviations from zero at $>$1$\sigma$ in any transition; and (iii) the `composite residual spectrum' (CRS) -- the average of the (1$\sigma$) normalised residuals when registered on the same velocity axis \citep{Malec:2010:1541} -- shows no significant evidence for unmodelled structure common to many transitions. Criterion (i) allows a somewhat larger $\chi^2_\nu$ than allowed by \citet{Murphy:2016:2461} who analysed only \ion{Zn}{ii} and \ion{Cr}{ii} transitions, which have similar strengths and are more easily modelled. However, $\chi^2_\nu$ values of $\approx$1.5 were achieved in \citet{Molaro:2013:A68} and \citet{Evans:2014:128} where all available transitions, with widely varying strengths, were considered. Similar to \citet{Evans:2014:128}, we also find that some parts of the extracted Subaru/HDS exposures have slightly (factors of 1.1--1.3) larger root-mean-square (RMS) flux variations than expected in unabsorbed regions. In our experience, this appears to be an unavoidable characteristic of {\sc iraf}-reduced echelle spectra. Rather than artificially increasing the error arrays to match the RMS flux variations around our transitions of interest, we instead allow a $\chi^2_\nu$ threshold of 1.5.

The assumption of turbulent broadening means that the $b$-parameters of corresponding components in different ionic species are tied to take the same value throughout the model trialling and $\chi^2$ minimisation processes. Alternatively, the assumption of purely thermal broadening could have been made. However, it is important to realise that we do not aim to accurately derive individual components' physical properties but, instead, aim to measure \daa\ which is a single parameter constrained by all components of all transitions. It is therefore most important to fit all the statistically significant structure in the absorption profiles of all transitions (using the same velocity structure). This is the primary aim of the fitting approach described above, i.e.\ maximising the number of fitted components by minimising $\chi^2_\nu$ as the main model selection criterion. Indeed, \citet{Murphy:2003:609}, \citet{King:2012:3370} and
\citet{Evans:2014:128} have previously found the effect of assuming different broadening mechanisms on \daa\ to be unimportant.

The laboratory wavelengths, oscillator strengths, isotopic and hyperfine structures used here were recently reviewed in \citet{Murphy:2014:388}. We only use the transitions recommended there -- i.e.\ only those with laboratory wavelength uncertainties corresponding to $<$20\,\ms. This includes the \tran{Ca}{ii}{$\lambda$3934/3969} (i.e.\ K and H) doublet in the $\zab=0.576$ absorber towards J0120$+$2133. While this is the first time these transitions have been used for a MM analysis to our knowledge, the \ion{Ca}{ii} absorption is quite weak in this case. We use the terrestrial isotopic abundances of all elements in our models, and we consider the systematic effects this may introduce in \Sref{sss:iso}.

\subsection{Fits to individual absorption systems}\label{ss:fits}

\begin{table*}
\begin{center}
  \caption{Main results for each absorption system. The quasar name and absorption redshift (\zab) correspond to the absorption profile fits plotted in Figs.\ \ref{f:J0120_0.5764}--\ref{f:J1944_1.7384}. The other columns provide the best-fit value of \daa\ (assuming terrestrial isotopic abundances; see \Sref{sss:iso} for discussion), its 1$\sigma$ statistical uncertainty ($\sigma_{\rm stat}$), the components of the systematic error budget (``LRD'' for long-range distortions, \Sref{sss:long}; ``IOD'' for intra-order distortions, \Sref{sss:intra}; ``Redisp.'' for redispersion errors, \Sref{sss:redisp}) and total systematic uncertainty ($\sigma_{\rm sys}$, the quadrature sum of the components, \Sref{ss:mainres}), and the final $\chi^2$ per degree of freedom, $\chi^2_\nu$, for each absorption system. For the $\zab=1.325$ absorber towards J0120$+$2133, an additional systematic error component of 1.44\,ppm, due to absorption profile modelling uncertainties (\Sref{sss:moderr}), is included in the quadrature sum to estimate $\sigma_{\rm sys}$. The sign of the LRD uncertainty indicates the sense in which \daa\ changes when a positive additional distortion slope is introduced into the spectrum. The final entry for J0120$+$2133 provides the weighted mean \daa\ and its uncertainties for that line of sight, taking into account the common-mode LRD uncertainties (see \Sref{ss:mainres}). The modelling uncertainty for the $\zab=1.325$ absorber contributes 0.95\,ppm to the total $\sigma_{\rm sys}$ for the J0120$+$2133 sight-line.}
\label{t:res}
\begin{tabular}{:l;c;c;c;c;c;c;c;c}\hline
Quasar & \zab & \daa & $\sigma_{\rm stat}$ & \multicolumn{4}{c}{Systematic errors [ppm]} & $\chi^2_\nu$ \\
       &      & [ppm] & [ppm]             & LRD      & IOD      & Redisp.  & $\sigma_{\rm sys}$ & \\\hline
J0120$+$2133  & 0.576   & $ -9.12 $ & 39.80 & $ -1.32 $ &  0.19 &  3.38 &  3.64 & 1.05 \\ 
J0120$+$2133  & 0.729   & $  0.73 $ &  6.17 & $  1.67 $ &  0.20 &  0.55 &  1.77 & 1.16 \\ 
J0120$+$2133  & 1.048   & $  5.47 $ & 18.26 & $  3.54 $ &  0.14 &  1.90 &  4.02 & 1.26 \\ 
J0120$+$2133  & 1.325   & $  2.60 $ &  3.45 & $  1.60 $ &  0.12 &  1.02 &  2.38 & 1.45 \\ 
J0120$+$2133  & 1.343   & $  8.36 $ & 11.82 & $  2.65 $ &  0.20 &  1.00 &  2.84 & 1.42 \\ 
J0120$+$2133  & Ave.    & $  2.53 $ &  2.87 & $  1.73 $ &  0.10 &  0.68 &  2.10 & ---  \\ 
J1944$+$7705  & 1.738   & $ 12.70 $ & 16.13 & $  0.59 $ &  0.12 &  1.34 &  1.47 & 1.22 \\ 
\hline
\end{tabular}

\end{center}
\end{table*}

In this section we plot the spectrum and fit for each absorption system and discuss any relevant complexities or difficulties encountered. The final best-fit parameters and uncertainties output from {\sc vpfit} are provided electronically in \citet{Murphy:2017:alphaSubaru} so that readers can fully reproduce the results presented here. We summarise the main results for each absorption system in \Tref{t:res}, including the measured value of \daa, its 1$\sigma$ statistical uncertainty, and $\chi^2_\nu$ derived from the fit.

\subsubsection{$\zab=0.576$ towards J0120$+$2133}\label{sss:J0120_0.5764}

\begin{figure}
\begin{center}
\includegraphics[width=1.0\columnwidth]{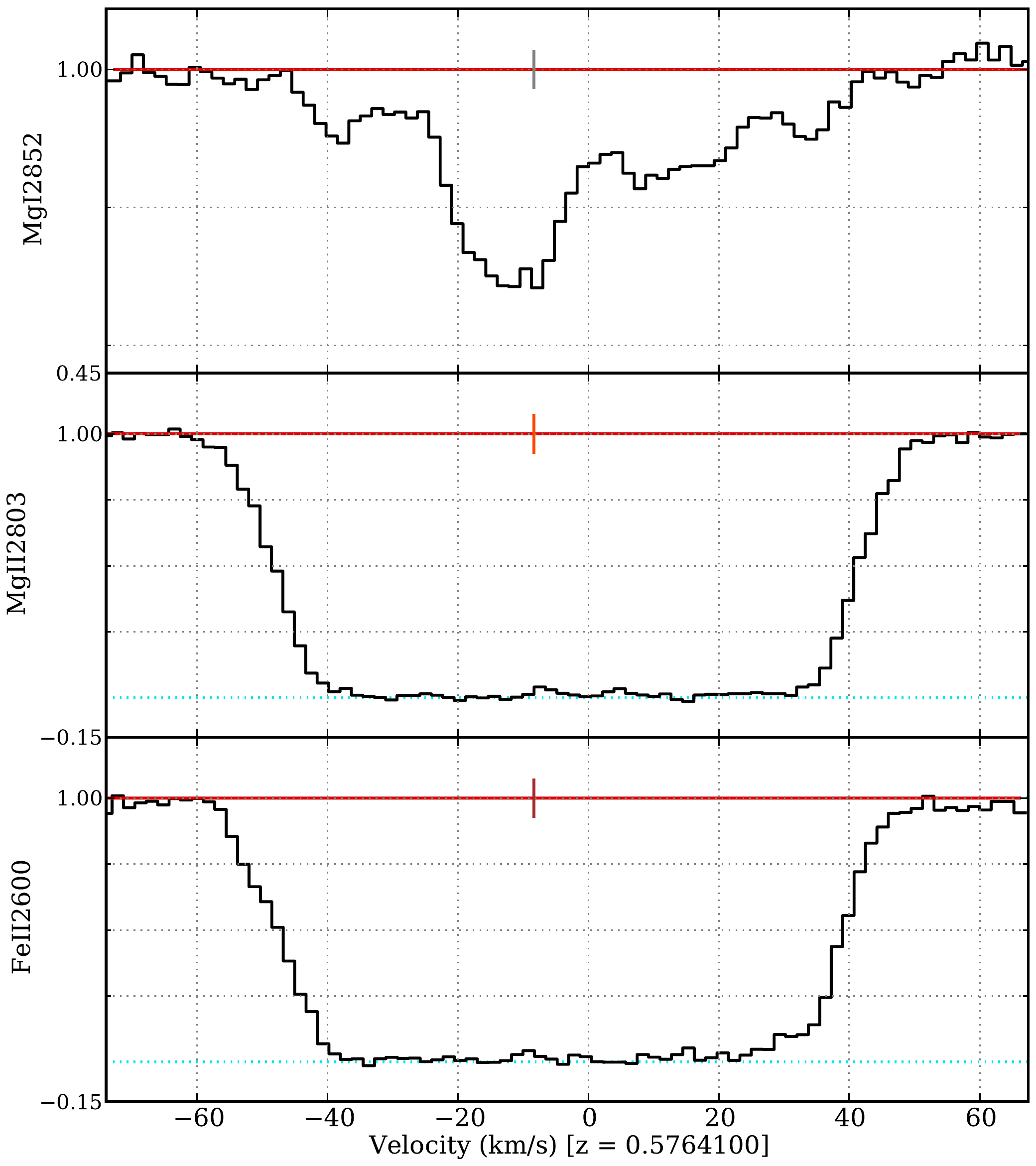}\vspace{-1em}
\caption{Examples of the heavily saturated \ion{Mg}{ii} and \ion{Fe}{ii} lines, compared to the \tran{Mg}{i}{2852} transition, in the $\zab=0.576$ absorber towards J0120$+$2133 (see \Sref{sss:J0120_0.5764}). The continuum-normalised flux spectra (black histograms) are plotted on a common velocity axis, relative to redshift $z=0.57641$, for all transitions. Note that the plotted range in normalised flux is different for the \ion{Mg}{i} (top panel) and saturated \ion{Mg}{ii} and \ion{Fe}{ii} transitions (lower panels). \tran{Mg}{i}{2852} is the strongest transition used to constrain \daa\ in this absorber (see \Fref{f:J0120_0.5764}); \ion{Mg}{ii} and \ion{Fe}{ii} were not used because no part of their profiles can be reliably modelled and compared with any part of \tran{Mg}{i}{2852}'s profile (or that of any other transition in \Fref{f:J0120_0.5764}).}
\label{f:J0120_0.5764_MgFe}
\end{center}
\end{figure}

The very common \tran{Mg}{ii}{$\lambda$2796/2803} doublet and the two strong \ion{Fe}{ii} transitions in the wavelength range of the spectrum ($\lambda\lambda$2586/2600) are heavily saturated at all velocities in this absorber. This is illustrated in \Fref{f:J0120_0.5764_MgFe} where \tran{Mg}{ii}{2803} and \tran{Fe}{ii}{2600} are compared with the next strongest (but much weaker) transition \tran{Mg}{i}{2852}. There is no common, unsaturated part of the \ion{Mg/Fe}{ii} and \ion{Mg}{i} profiles. Therefore, even in principle, including \ion{Mg}{ii} and \ion{Fe}{ii} in a joint profile fit with the unsaturated transitions in this absorber would not contribute important constraints on \daa. In practice, it is also very difficult to fit such uniformly saturated profiles when no unsaturated transitions of the same species are available to constrain the velocity structure and column densities. Therefore, we do not include the \ion{Mg}{ii} and \ion{Fe}{ii} transitions in our analysis of this absorber because they offer no practical constraints on \daa.

\begin{figure}
\begin{center}
\includegraphics[width=1.0\columnwidth]{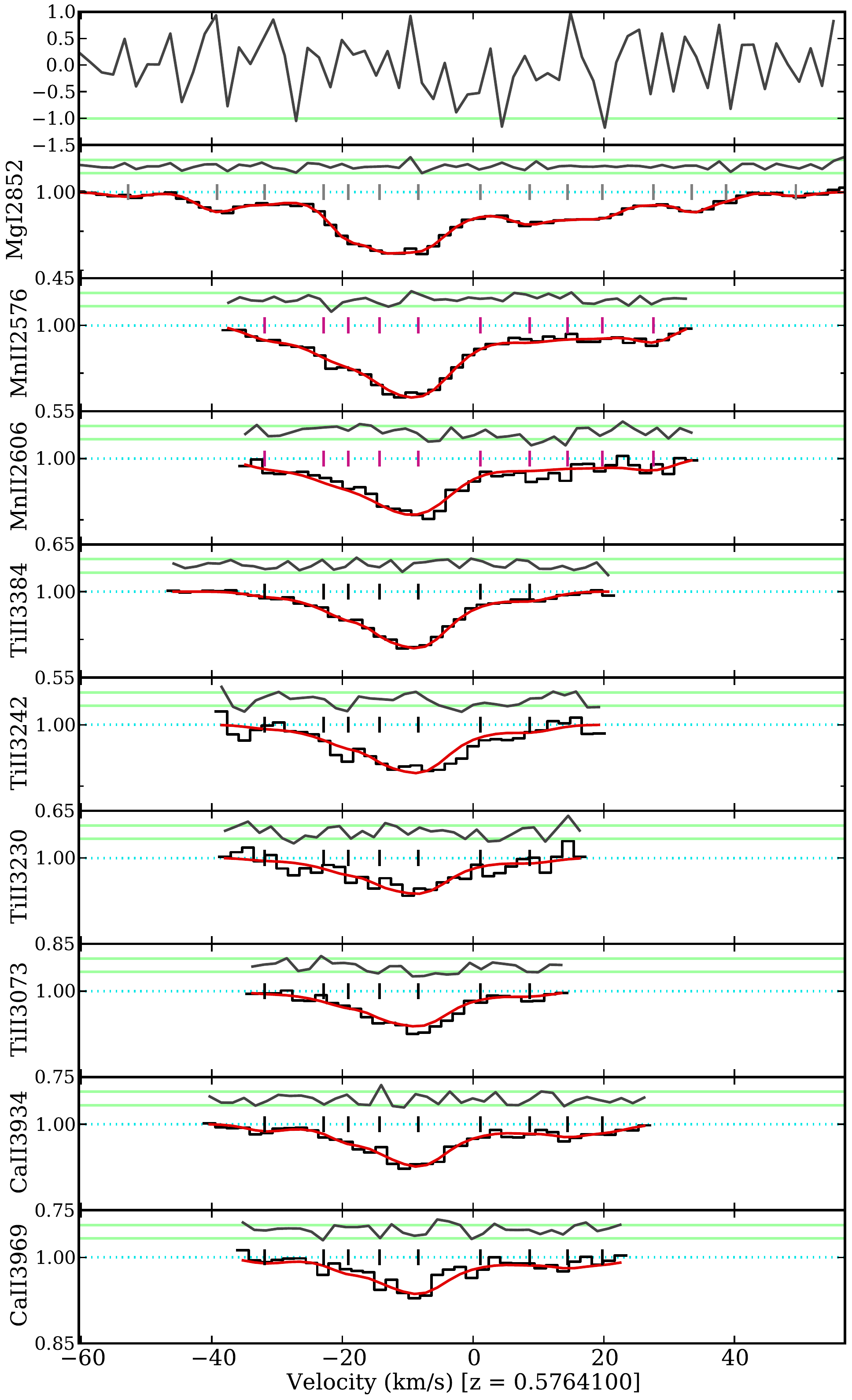}\vspace{-1em}
\caption{Fit to the Subaru/HDS spectrum of the $\zab=0.576$ absorber towards J0120$+$2133 (see \Sref{sss:J0120_0.5764}). The lower nine panels show the continuum-normalised flux density versus velocity (relative to the redshift in brackets) for the transitions fitted to measure \daa\ which are labelled on the vertical axis. Note that the plotted range in normalised flux is different for each transition and depends the maximum absorption depth. The best fit (red solid line) overlays the spectrum (black histogram) and comprises Voigt profile components at velocities indicated by the vertical ticks; the isotopic and hyperfine structures of each component are represented by a single tick. The (1$\sigma$) normalised residuals between the spectrum and fit are plotted above each transition, with $\pm$1$\sigma$ flux uncertainties marked by the horizontal solid lines. The composite residual spectrum (CRS; black line) is shown in the top panel, in units of standard deviations. The CRS is the average of the normalised residuals from the transitions registered to each other on the velocity axis shown; when all statistically significant structure in the absorption profile is reflected by the model fit, the CRS is expected to show no significant structure.}
\label{f:J0120_0.5764}
\end{center}
\end{figure}

The fit to the 9 transitions used for the MM analysis of this absorber are plotted in \Fref{f:J0120_0.5764}. The strongest transition used to constrain \daa\ is \tran{Mg}{i}{2852} which extends over $\sim$110\,\kms. However, most of the absorption is confined to one main, $\approx$25\,\kms-wide spectral feature centred near $-10$\,\kms. This provides most of the constraints on \daa. However, this feature is not especially narrow or sharp; it is instead rather `smooth', which means that we cannot expect a very precise measurement of \daa\ in this case. Compounding this is the fact that only 4 of the transitions absorb $\ga$20\% of the continuum flux (\tran{Mg}{i}{2852}, \tran{Mn}{ii}{$\lambda$2576/2606} and \tran{Ti}{ii}{3384}); the other transitions will only very weakly constrain \daa. Indeed, \Tref{t:res} shows that the statistical uncertainty alone is 40\,ppm, much larger than the mean value from the other absorbers (11\,ppm).

Several transitions from the $\zab=0.729$ absorber (see \Sref{sss:J0120_0.7291}) below) blend with those in this absorber: \tran{Ti}{ii}{3067} cannot be fitted reliably in this absorber because it is strongly blended with \tran{Mg}{ii}{2796} at $\zab=0.729$; the fit to the red wing of \tran{Ti}{ii}{3073} is truncated to avoid blending with \tran{Mg}{ii}{2803} at $\zab=0.729$; and the absorption bluewards of $\approx-28$\,\kms\ near \tran{Mg}{i}{2852} may be partially due to \tran{Fe}{ii}{2600} at $\zab=0.729$. In the latter case, none of the fitted \tran{Mg}{i}{2852} velocity components that are potentially blended have counterparts in any other transition, so they will not affect \daa; we simply fit them as \tran{Mg}{i}{2852} for convenience.

The fit in \Fref{f:J0120_0.5764} satisfies all the selection criteria discussed in \Sref{ss:genfit}: there are no significant, long-range, coherent deviations in the normalised residuals in any transition; there is no evidence in the CRS (top panel in \Fref{f:J0120_0.5764}) for additional, unmodelled velocity structure; and \Tref{t:res} shows that the final $\chi^2_\nu$ for this fit is less than the threshold we impose of 1.5. Therefore, this absorber is accepted for our statistical analysis in \Sref{ss:mainres} below.

\subsubsection{$\zab=0.729$ towards J0120$+$2133}\label{sss:J0120_0.7291}

\begin{figure}
\begin{center}
\includegraphics[width=1.0\columnwidth]{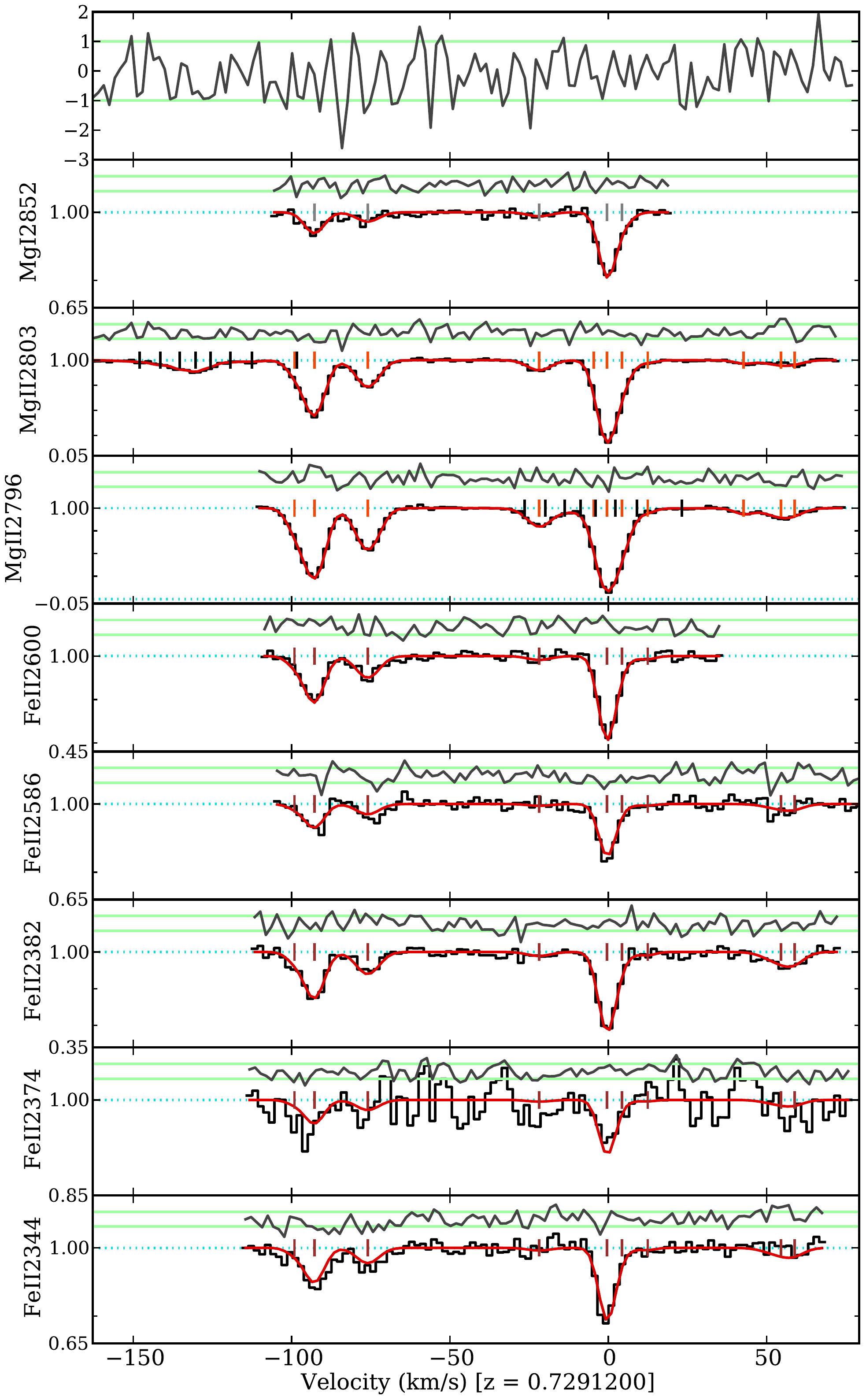}\vspace{-1em}
\caption{Same as \Fref{f:J0120_0.5764} but for the $\zab=0.729$ absorber towards J0120$+$2133 (see \Sref{sss:J0120_0.7291}).}
\label{f:J0120_0.7291}
\end{center}
\end{figure}

The 8 transitions used to constrain \daa\ in this absorber, plotted in \Fref{f:J0120_0.7291}, are the most common in MM analysis because they are the strongest \ion{Mg}{i/ii} and \ion{Fe}{ii} transitions available \citep[of those with precisely known laboratory wavelengths in][]{Murphy:2016:2461}. The \ion{Fe}{ii} transitions all have large, positive $q$ coefficients, whereas the \ion{Mg}{i/ii} transitions are much less sensitive to $\alpha$ variation, making their combination very important for constraining \daa. The absorption in this system is spread over 185\,\kms\ with 3 main spectral features, but the feature at 0\,\kms\ is the strongest in all transitions and is relatively narrow/sharp, so it constrains \daa\ most strongly.

As noted in \Sref{sss:J0120_0.5764} above, the \ion{Mg}{ii} doublet transitions are both blended with \ion{Ti}{ii} transitions in the $\zab=0.576$ absorber. These \ion{Ti}{ii} blends are shown as black tick marks in \Fref{f:J0120_0.7291} and their velocity structure is most clearly seen in the \tran{Mg}{ii}{2803} panel. The blending occurs at different places in the velocity structure of the two \ion{Mg}{ii} transitions in such a way that the entire \ion{Ti}{ii} velocity structure is independently constrained (i.e.\ the column density, $b$-parameter and redshift of each component is fully constrained, without blending, in at least one \ion{Ti}{ii} transition). We therefore fitted these two \ion{Ti}{ii} transitions ($\lambda$3067 and $\lambda$3073) with a similar velocity structure as that in \Sref{sss:J0120_0.5764} but with the component parameters allowed to vary freely (\daa\ was fixed to zero; allowing it to vary freely does not affect \daa\ in the absorber of interest at $\zab=0.729$). This allows the most conservative incorporation of their uncertainties into that of \daa\ via the off-diagonal terms of the covariance matrix in the $\chi^2$ minimisation process. Unsurprisingly, these parameters have larger uncertainties than, but are consistent with, those from the \ion{Ti}{ii} fit in the $\zab=0.576$ itself (in \Sref{sss:J0120_0.5764}) where the velocity structure is determined jointly by all transitions. Finally, the reddest spectral feature is not fitted in \tran{Fe}{ii}{2600} due to the blend with \tran{Mg}{i}{2852} in the $\zab=0.576$ absorber already discussed (\Sref{sss:J0120_0.5764}).

The fit shown in \Fref{f:J0120_0.7291} passes our selection criteria, showing no concerning structures in the individual transitions' residuals, nor the CRS, and the final $\chi^2_\nu$ is less than 1.5. This absorber is therefore included in our statistical analysis.

\subsubsection{$\zab=1.048$ towards J0120$+$2133}\label{sss:J0120_1.0480}

\begin{figure}
\begin{center}
\includegraphics[width=1.0\columnwidth]{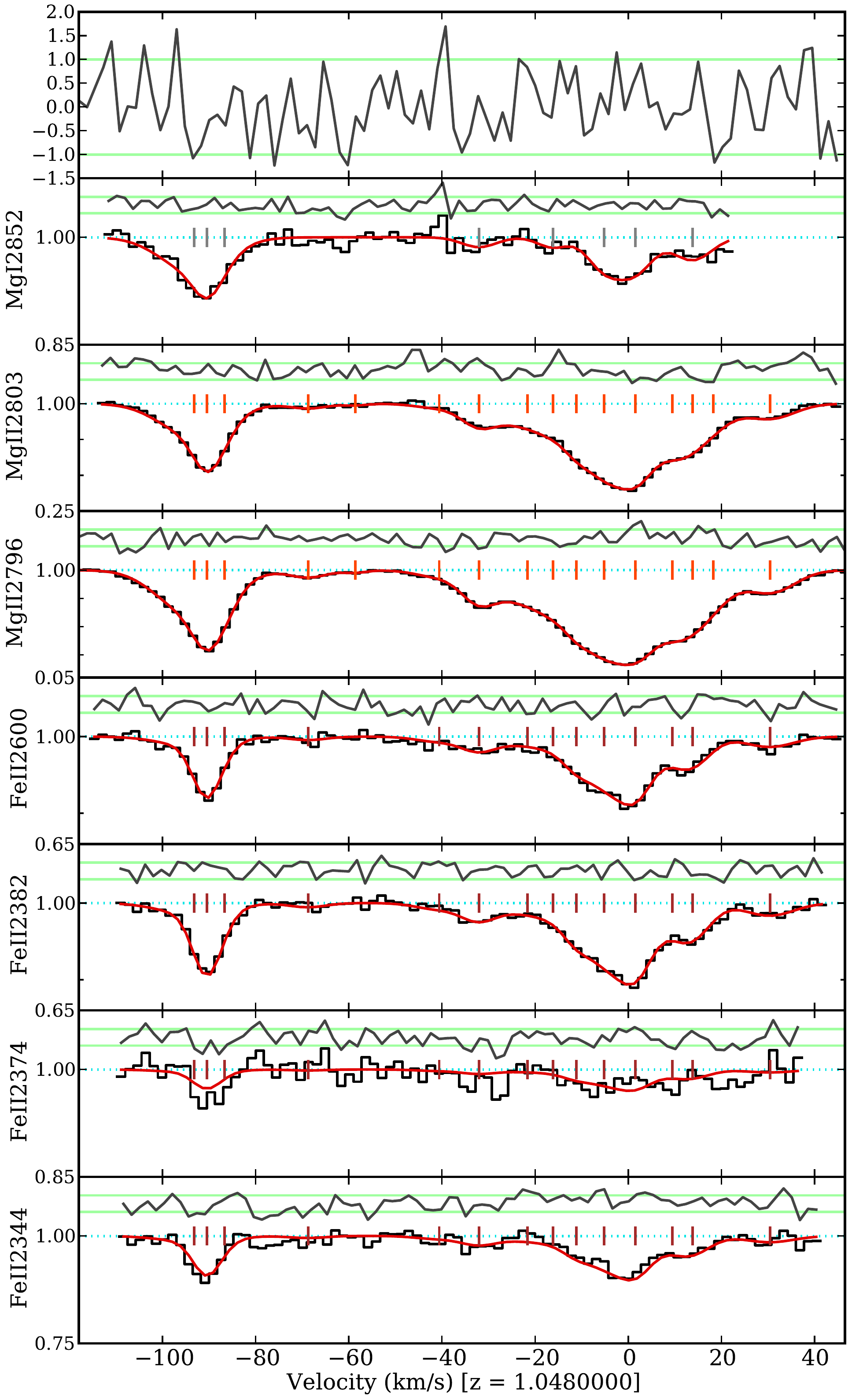}\vspace{-1em}
\caption{Same as \Fref{f:J0120_0.5764} but for the $\zab=1.048$ absorber towards J0120$+$2133 (see \Sref{sss:J0120_1.0480}).}
\label{f:J0120_1.0480}
\end{center}
\end{figure}

\daa\ in this absorber was constrained using \tran{Mg}{i}{2852}, the \ion{Mg}{ii} doublet and 4 of the 5 strongest \ion{Fe}{ii} transitions at $\lambda>2300$\,\AA, as shown in \Fref{f:J0120_1.0480}. The \tran{Fe}{ii}{2586} transition was excluded because it was strongly affected by artefacts in the spectrum, likely from scattered light within the spectrograph. The absorption occurs in two main features, centred around 0 and $-92$\,\kms, and spread over 150\,\kms\ in total. While the feature at $-92$\,\kms\ is relatively sharp, it is weaker and less prominent in \ion{Mg}{i} and \ion{Fe}{ii}, while the feature around 0\,\kms\ is relatively smooth. Therefore, we do not expect strong constraints on \daa\ in this absorber and, indeed, \Tref{t:res} shows this to be the case, with a 1$\sigma$ statistical uncertainty of $18$\,ppm.

We did not identify any blends in this system, so the fit in \Fref{f:J0120_1.0480} is relatively uncomplicated. Some components in the strongest transitions (\ion{Mg}{ii}) were not statistically justified in \tran{Mg}{i}{2852} or, in the case of one component near $-60$\,\kms, the \ion{Fe}{ii} transitions. However, these were the weakest components and omitting them from the \tran{Mg}{i}{2852} and \ion{Fe}{ii} profiles is consistent with our fitting approach (and does not affect \daa). The residuals, CRS and $\chi^2_\nu$ for this fit all pass our selection criteria, so this absorber is included in our main results.

\subsubsection{$\zab=1.325$ towards J0120$+$2133}\label{sss:J0120_1.3254}

\begin{figure}
\begin{center}
\includegraphics[width=1.0\columnwidth]{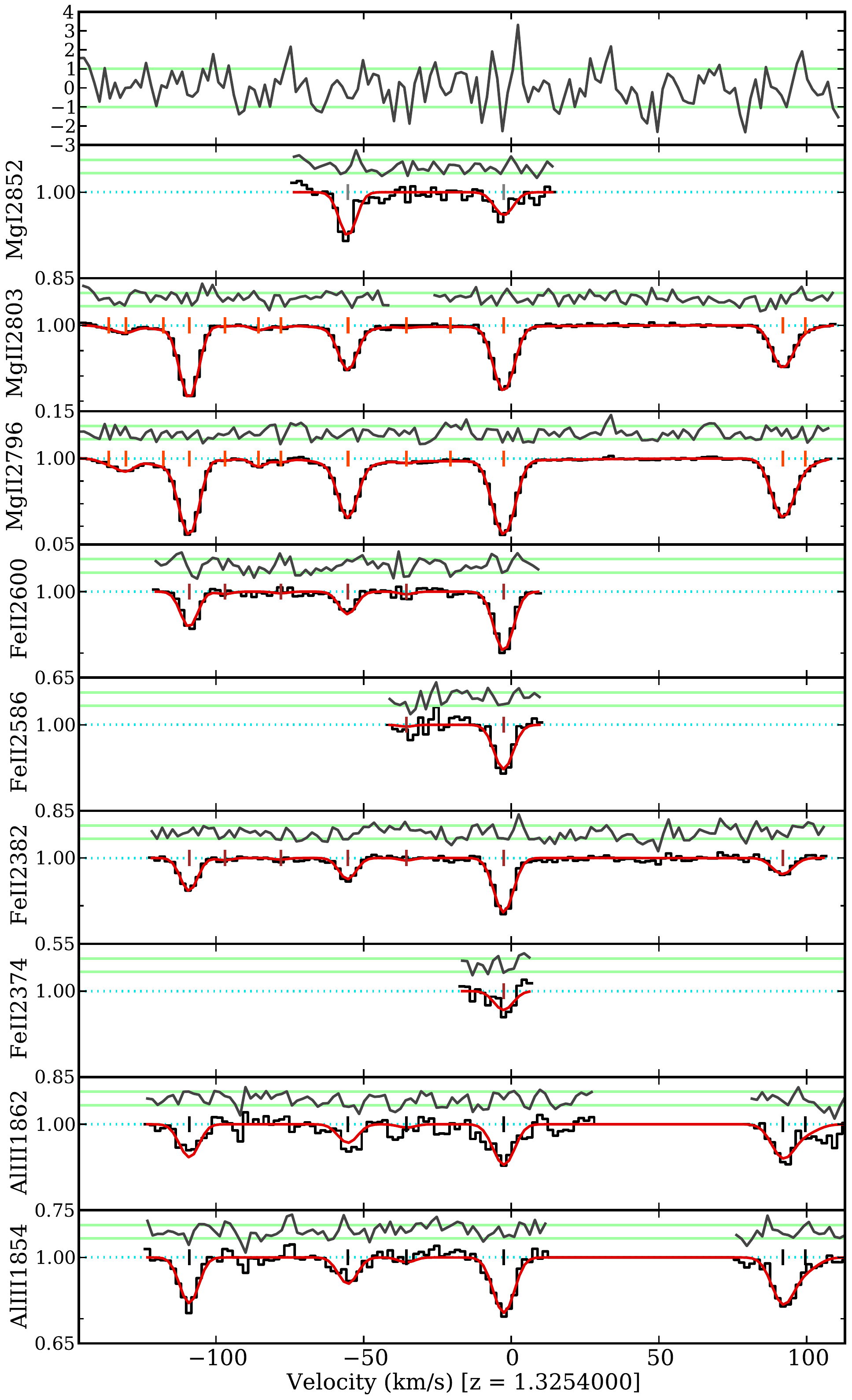}\vspace{-1em}
\caption{Same as \Fref{f:J0120_0.5764} but for the $\zab=1.325$ absorber towards J0120$+$2133 (see \Sref{sss:J0120_1.3254}).}
\label{f:J0120_1.3254}
\end{center}
\end{figure}

The common combination of \ion{Mg}{i/ii} and \ion{Fe}{ii} transitions are used in this absorber, together with the \ion{Al}{iii} doublet. \Fref{f:J0120_1.3254} shows that the \ion{Fe}{ii} and \ion{Al}{iii} absorption is relatively weak, so only the data sections with detectable absorption were included in the fit for those transitions. Nevertheless, the absorption comprises 4 sharp, well-separated spectral features, spread over $\approx$275\,\kms, so we can expect \daa\ to be well constrained. \Tref{t:res} shows that the 1$\sigma$ statistical error is indeed relatively small, just $3.5$\,ppm which is the smallest of any of the absorbers studied here. The \tran{Fe}{ii}{2344} transition fell in the gap between the two Subaru/HDS CCD chips and so could not be included in the fit. Note that the \ion{Mg}{ii} profiles required 2 components -- one narrow and the other broad -- to be fitted in the spectral feature near $-60$\,\kms. However, these components are at very similar velocities (their tick marks are not easily distinguishable in \Fref{f:J0120_1.3254}), though the broader component is apparent from the shape of the \tran{Mg}{ii}{2796} profile. The broader component is weak in \ion{Mg}{ii} and, given that the spectral feature is very weak in the other species, it is not surprising that it was rejected from the fit by {\sc vpfit} (i.e.\ it is not statistically required). Alternative profile models for this absorption system are considered in detail in \Sref{sss:moderr}.

Some small sections of \tran{Mg}{ii}{2803} (at $-35$\,\kms) and, particularly, the \ion{Al}{iii} doublet transitions ($\sim$20--70\,\kms) were excised from the fit due to artefacts evident in some or all contributing exposures (found by comparing the exposures in {\sc uves\_popler}). However, these did not seem to affect the main spectral features. We expect weak telluric absorption lines in the region of the spectrum where the \ion{Mg}{i/ii} transitions fall and, upon inspection, some features are evident near these transitions. However, we do not detect any inconsistencies between these transitions when constructing the fit to this absorber. Nor do we observe any evidence of these in the normalised residuals or CRS in \Fref{f:J0120_1.3254}, and the final $\chi^2_\nu$ is $1.45$, slightly below our threshold of 1.5. We therefore proceed with using \ion{Mg}{i/ii} to constrain \daa\ and include this absorber in our statistical analysis.

\subsubsection{$\zab=1.343$ towards J0120$+$2133}\label{sss:J0120_1.3430}

\begin{figure}
\begin{center}
\includegraphics[width=1.0\columnwidth]{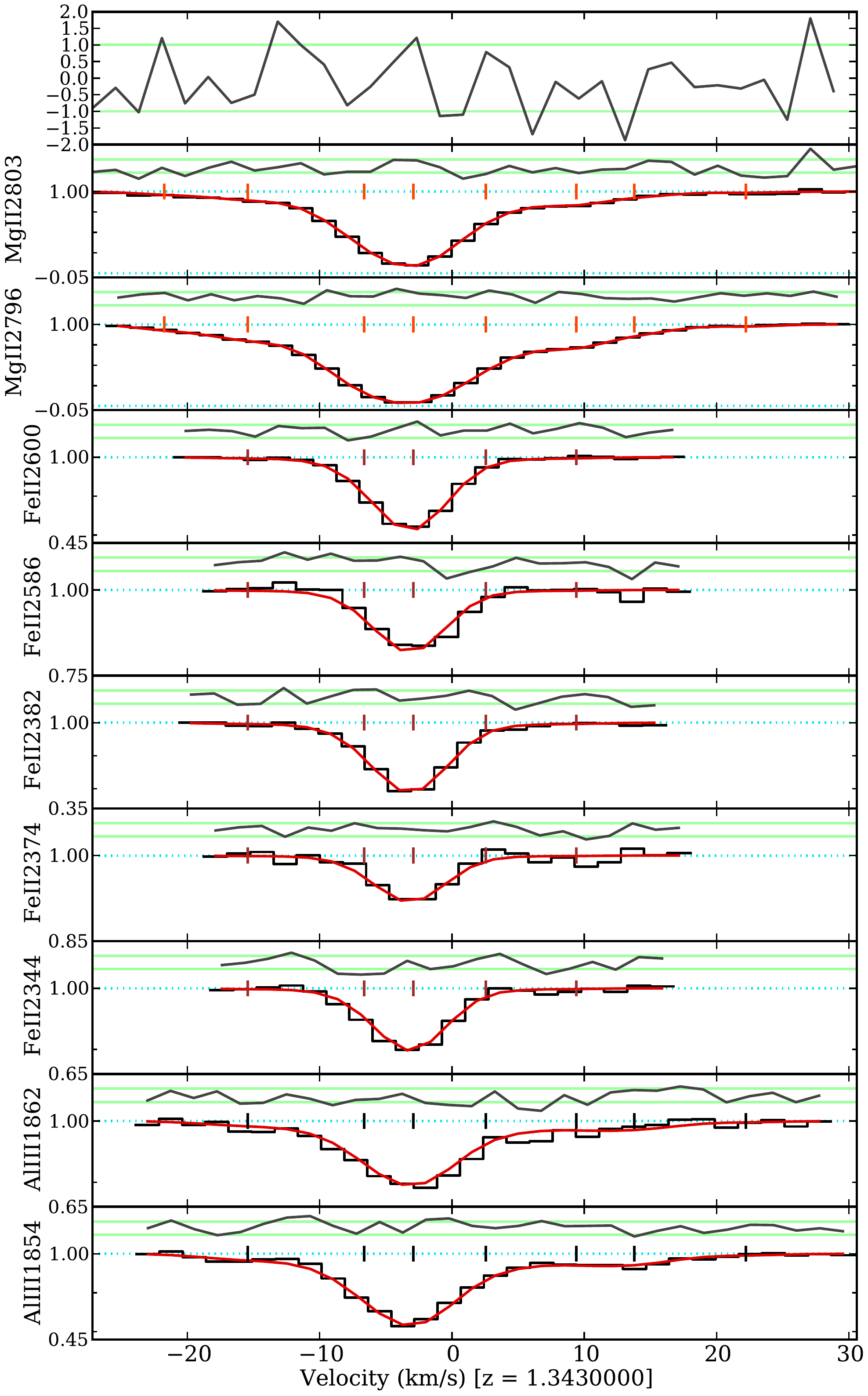}\vspace{-1em}
\caption{Same as \Fref{f:J0120_0.5764} but for the $\zab=1.343$ absorber towards J0120$+$2133 (see \Sref{sss:J0120_1.3430}).}
\label{f:J0120_1.3430}
\end{center}
\end{figure}

The 9 transitions used to constrain \daa\ in this absorber are the same as those in the nearby $\zab=1.325$ system above, as shown in \Fref{f:J0120_1.3430}. However, in this case there is only one, slightly broader, main spectral feature. We therefore expect a somewhat larger uncertainty in \daa\ compared to the $\zab=1.325$ system, and \Tref{t:res} demonstrates that is indeed the case. Significant \ion{Mg}{ii} absorption is detected bluewards of the main feature (to $\approx-150$\,\kms) but the corresponding absorption in even the strongest \ion{Fe}{ii} transition is only barely detected, so it will not constrain \daa. We therefore restrict the fit to the velocity range $-28$--30\,\kms\ for simplicity.

\Fref{f:J0120_1.3430} shows no significant complications with the fit to this absorber. As expected, the weakest components of the strong \ion{Mg}{ii} doublet are not statistically necessary in the weaker \ion{Fe}{ii} and \ion{Al}{iii} transitions, so are not fitted. The residuals, CRS and $\chi^2_\nu$ pass our selection criteria, so this absorber is included in our main results below.

\subsubsection{$\zab=1.119$ towards J1944$+$7705}\label{sss:J1944_1.1190}

We intended to constrain \daa\ using this absorber, in which existing Keck/HIRES spectra showed strong \ion{Mg}{ii} absorption and weak \tran{Fe}{ii}{2344}, $\lambda$2382 and $\lambda$2600 lines. However, severe data artefacts preclude detection of \tran{Fe}{ii}{2600} and smaller artefacts made any fit to \tran{Fe}{ii}{2344} very uncertain and likely spurious. The remaining \ion{Fe}{ii} transition, $\lambda$2382, has lower \SN\ than planned which, on its own, does not provide an accurate or precise velocity measurement when fitted together with the almost saturated \ion{Mg}{ii} lines. Therefore, we did not attempt to constrain \daa\ with this absorber.

\subsubsection{$\zab=1.738$ towards J1944$+$7705}\label{sss:J1944_1.7384}

\begin{figure}
\begin{center}
\includegraphics[width=1.0\columnwidth]{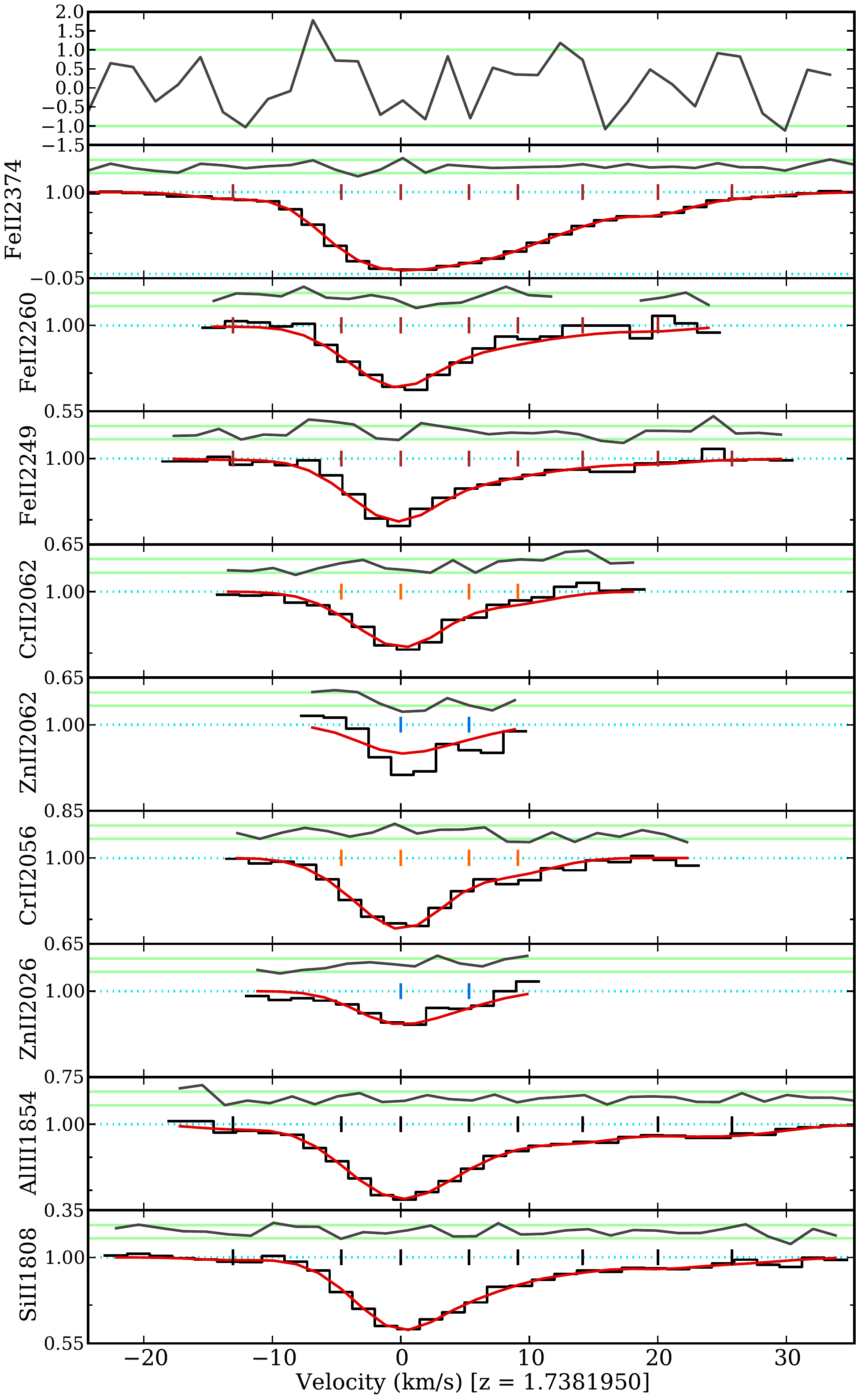}\vspace{-1em}
\caption{Same as \Fref{f:J1944_1.7384} but for the $\zab=1.738$ absorber towards J1944$+$7705 (see \Sref{sss:J1944_1.7384}).}
\label{f:J1944_1.7384}
\end{center}
\end{figure}

This is the highest-redshift absorber in this work. Consequently, a wider range of different ionic species is available for constraining \daa, as shown in \Fref{f:J1944_1.7384}, and the transitions have a very large range of $q$ coefficients \citep[e.g.\ $-1280$\,cm$^{-1}$ for \tran{Cr}{ii}{2062} through to $2470$\,cm$^{-1}$ for \tran{Zn}{ii}{2026};][]{Murphy:2014:388}, potentially constraining \daa\ very strongly. On the other hand, some of these transitions are relatively weak and there is only one main spectral feature in this absorber, spanning $\approx$55\,\kms, which is fairly smooth. Taken together, these factors lead us to expect a moderate constraint on \daa\ and, indeed, \Tref{t:res} shows this to be the case; the 1$\sigma$ statistical uncertainty is 16\,ppm, approximately the mean from all the absorbers studied here.

The fit to the two very weak \ion{Zn}{ii} transitions only requires the strongest component and one flanking component, located at $\approx$6\,\kms, but not the other flanking component at $\approx-5$\,\kms. In other species, these two flanking components have similar strength, but \ion{Zn}{ii} is so weak that it is not unexpected, or important for \daa, that one component may not be statistically required to fit the \ion{Zn}{ii} profiles. The fit to the absorber in \Fref{f:J1944_1.7384} does not show any problems in the residuals, the CRS or the final $\chi^2_\nu$ of 1.2. It is therefore accepted into our statistical analysis below.

\section{Results}\label{s:res}

\subsection{\boldmath{$\daa$} measurements and statistical uncertainties}\label{ss:dares}

The \daa\ results, with statistical uncertainties, from the fits to the 6 absorbers in Figs.\ \ref{f:J0120_0.5764}--\ref{f:J1944_1.7384} are shown in \Tref{t:res}. Aside from the $\zab=1.119$ towards J1944$+$7705 which could not be fitted (\Sref{sss:J1944_1.1190}), all 6 absorbers in \Tref{t:res} pass our selection criteria. The \daa\ values range from $-9$ to 13\,ppm, with an average statistical uncertainty of 16\,ppm. However, the 1$\sigma$ statistical uncertainties have a large range, 3.5--40\,ppm. Nevertheless, all 6 results are individually consistent with $\daa=0$, and also consistent with each other, within their statistical uncertainties.

The weighted mean \daa\ of the 6 results in \Tref{t:res}, using only the statistical uncertainties, is $\left<\daa\right>_{\rm w}=2.9\pm2.8$\,ppm. That is, even without considering systematic uncertainties, there is no evidence for a varying $\alpha$ from these results. The statistical precision here, $\approx$3\,ppm, is comparable to that in other recent MM analyses using small samples ($\la$10) of absorbers \citep[e.g.][]{Molaro:2013:A68,Evans:2014:128} but somewhat larger than that obtained from the large Keck and VLT samples \citep[1.2\,ppm][]{Murphy:2004:131,King:2012:3370}, the very high-\SN\ spectrum of \citet{Kotus:2017:3679} and the ensemble of 11 \ion{Zn/Cr}{ii}-only measurements of \citet{Murphy:2016:2461}. \daa\ has a significantly smaller statistical uncertainty in two of our absorbers, $\zab=0.729$ (6.2\,ppm) and especially $\zab=1.325$ (3.5\,ppm) towards J0120$+$2133, than in the others, so these dominate the weighted mean. Nevertheless, removing the $\zab=1.325$ measurement does not change the weighted mean substantially: $3.4\pm4.9$\,ppm.

The $\chi^2$ around the weighted mean \daa\ of all 6 absorbers (again using only the statistical uncertainties) is 0.83 which, for 5 degrees of freedom, is somewhat lower than expected (i.e.\ $\sim$5). However, the probability that $\chi^2$ should be this small or smaller, simply by random chance, is 2.5\% (equivalent to $\approx$2.1$\sigma$). While this is not highly statistically significant, it is still worth considering whether biases exist in our \daa\ values or whether the statistical uncertainties are overestimated. With our profile fitting approach of including as many velocity components as statistically supported by the data, and by fixing \daa\ to zero when establishing the fiducial velocity structures, it may be possible that our results are biased towards zero. However, we conducted thorough tests to ensure that {\sc vpfit} converged on the best-fit \daa: for every absorber, we (i) ran the $\chi^2$ minimisation many times over (at least 30 times for every absorber), starting each new run from the end point of the last, to test whether \daa\ was still changing appreciably between runs; and (ii) fixed \daa\ to the value in \Tref{t:res} plus (minus) half its statistical uncertainty and ran the $\chi^2$ minimisation several times, then allowed \daa\ to be a free parameter and repeated the test for convergence; in all cases \daa\ changed back towards the value in \Tref{t:res}, and did not move away from it as might be expected if the original value was not the best fit one. The other possibility -- overestimated uncertainties -- is unlikely because {\sc vpfit} has been tested thoroughly with simulated spectra \citep[e.g.][]{Murphy:2003:thesis,Murphy:2008:1611} and Monte Carlo Markov Chain analysis \citep{King:2009:864} and in all cases the statistical uncertainty is correctly estimated (the latter analysis indicated it may be overestimated by up to 10\%, not large enough for concern here). Instead, we may expect from the small scatter in \daa, relative to the statistical errors, that the latter dominate the error budget, i.e.\ that systematic errors are relatively small. We consider systematic uncertainties in the next section and, indeed, find this to be the case.

\subsection{Systematic uncertainties}\label{ss:syserr}

\subsubsection{Long-range distortions}\label{sss:long}

As discussed in \Sref{s:intro}, reliable measurements of \daa\ are likely only possible if the long-range distortions in the quasar are corrected. In \Sref{ss:scal} we used attached solar twin supercalibration exposures to correct all quasar exposures from the 2014 run (for both J0120$+$2133 and J1944$+$7705), so systematic errors will be mostly avoided for these exposures, with the main effect on \daa\ stemming from uncertainties in the distortion correction. However, we did not correct the 3 exposures of J0120$+$2133 from the 2012 run for lack of attached supercalibrations. Instead, non-attached asteroid supercalibration exposures indicate the distortions slopes may be $\sim$2.0\,\msnm, which is consistent with the lack of distortion found, at that precision level, between its 2012 and 2014 sub-spectra. This lack of correction and larger uncertainty in the distortion slope will, therefore, likely dominate the systematic errors for J0120$+$2133.

For J1944$+$7705, the slopes of the distortions in the individual supercalibration exposures was relatively small, $\la$0.3\,\msnm. However, note in \Fref{f:scal_Aug14} that the supercalibration slope varies by up to $\approx$0.5\,\msnm\ over the timescale of a quasar exposure ($\sim$1 hour). That is, we may expect the long-range distortions in the quasar exposures to differ from those in the supercalibrations by up to 0.5\,\msnm, but that this will vary in sign and magnitude in the 3 quasar exposures. To gauge the average effect on \daa, we therefore introduced an additional distortion slope of $0.5/\sqrt{3}=2.9$\,\msnm, with the same sign, to all 3 exposures in {\sc uves\_popler}, recombined the exposures, and re-ran the $\chi^2$ minimisation process for the $\zab=1.738$ absorber starting from the final fit and \daa\ value in \Tref{t:res}. This yielded a change in \daa\ by 0.59\,ppm, which we list as the systematic uncertainty from the long-range distortion corrections for this absorber in \Tref{t:res}.

For J0120$+$2133, the effect on \daa\ in each absorber from the uncertainty in the distortion slopes of the four 2014 exposures was determined in same way as for $\zab=1.738$ towards J1944$+$7705 above. As expected, this yielded only a small uncertainty of 0.1--0.5\,ppm for the 5 absorbers studied here. However, none of the three 2012 exposures were corrected for distortions and so may have all have the same $\sim$2.0\,\msnm\ distortion uncertainty. We therefore introduced a distortion of this magnitude and sign for these exposures and found that this caused much larger changes to \daa, 1.3--3.5\,ppm. In \Tref{t:res} we list the quadrature sum of the 2012 and 2014 effects on \daa\ as the long-range systematic uncertainty, though this is dominated by the former in all 5 absorbers. We also note that the change in \daa\ has the opposite sign for the $\zab=0.576$ absorber to the other 4 absorbers, and we indicated this with a negative sign in \Tref{t:res}; this is important to consider \Sref{ss:mainres} when forming the weighted mean \daa\ for the J0120$+$2133.

\subsubsection{Intra-order distortions}\label{sss:intra}

Intra-order distortions were first identified in Keck/HIRES quasar spectra by \citet{Griest:2010:158} and subsequently confirmed in VLT/UVES \citep{Whitmore:2010:89} and Subaru/HDS \citep{Evans:2014:128} supercalibration spectra of stars observed through iodine gas cells. Our solar twin and asteroid supercalibration exposures also reveal intra-order distortions with amplitudes up to $v_{\rm saw}\sim100$\,\ms\ with a simple ``saw-tooth'' structure (as a function of position along the echelle order) that is similar in all echelle orders. However, we find that the precise shape and amplitude of this saw-tooth pattern changes somewhat from exposure to exposure. By varying the position of the supercalibrator across the spectrograph slit, \citet{Whitmore:2015:446} found similar variations in the intra-order distortions. We therefore presume that the intra-order distortions in a quasar spectrum may somewhat differ from those in its corresponding supercalibration exposure, even if it was ``attached'' to the quasar one.

With this in mind, our approach to assessing the effect of intra-order distortions on \daa\ is to introduce a simple model of the distortions into our quasar spectra. We follow many recent works \citep[e.g.][]{Malec:2010:1541,Molaro:2013:A68,Bagdonaite:2014:10,Evans:2014:128,Murphy:2016:2461,Kotus:2017:3679} and adopt a basic saw-tooth model with a velocity shift amplitude of $+v_{\rm saw}$ at the echelle order centre with a linear fall-off to $-v_{\rm saw}$ at the order edges. This distortion is applied to each echelle order, of each quasar exposure, and these are then combined with {\sc uves\_popler} in the usual way. Using this modified spectrum, \daa\ is determined for each absorber using its fiducial, fitted velocity structure and starting the $\chi^2$ minimisation from the \daa\ value in \Tref{t:res}. For all 6 absorbers we find the systematic effect from this model of the intra-order distortions to be $\le0.2$\,ppm, which is much smaller than the long-range distortion uncertainties. This is similar to other recent MM analyses where all available transitions, at widely-separated wavelengths, were utilised; this has the effect of randomising the velocity shifts between transitions and, therefore, reducing the systematic effect on \daa\ in an individual absorber. Given that our saw-tooth model may not reflect the actual shape of the intra-order distortions in our quasar spectra, we adopt the mean effect on \daa\ across all 6 absorbers, 0.12\,ppm, as the minimum systematic uncertainty for any individual absorber. The results of this process are shown in \Tref{t:res} under the ``IOD'' column.

\subsubsection{Spectral redispersion}\label{sss:redisp}

The process of redispersing several quasar exposures onto a common grid, so that they may be combined, introduces correlations between neighbouring pixels (their fluxes and flux uncertainties) and, therefore, small distortions in the absorption line shapes. This can introduce small shifts between different transitions which will scale inversely with \SN\ and the number of contributing exposures. The magnitude and sign of the shift will also depend on the absorption line profile shapes and the particular alignments, or ``phases'', between the pixel grids of the contributing exposures and the common, redispersed grid; the effect is effectively random from from transition to transition. How these shifts affect \daa\ will further depend on which, and how many, transitions are fitted, i.e.\ the distribution of $q$ coefficients available.

To assess the systematic uncertainties on \daa\ from these redispersion effects, we follow the same approach as \citet{Murphy:2016:2461} and generate 8 alternative versions of each quasar spectrum, each with slightly different dispersion (pixel size) to the original combined spectrum: $\pm$0.01, $\pm$0.02, $\pm$0.03 and $\pm$0.04\,\kms. This has the same effect as changing the pixel phases of each contributing exposure relative to those in the redispersed grid. For each absorber, \daa\ is determined from each of the 8 spectra in the same way as the original result, using the same fiducial velocity structure and starting the $\chi^2$ minimisation process from the \daa\ value in \Tref{t:res}. As expected \citep[and found by ][]{Murphy:2016:2461}, the RMS of these 8 results and the original result, $\sigma_{\rm disp}$, is correlated with the statistical uncertainty, $\sigma_{\rm stat}$, across the 6 absorbers, with a linear least-squares fit yielding $\sigma_{\rm disp}[{\rm ppm}]\approx 0.06 + 0.08 \times \sigma_{\rm stat}[{\rm ppm}]$. Given that we only use 9 different measurements, there is a risk that $\sigma_{\rm disp}$ will be underestimated in an individual absorber. We therefore assign a minimum systematic uncertainty for redispersion effects to an individual absorber according to this linear relationship.

The results of this process are shown in \Tref{t:res} under the ``Redisp.'' column. Although the redispersion uncertainties are smaller than the long-range distortion uncertainties in the five J0120$+$2133 absorbers (in which the uncorrected 2012 exposures contribute the main uncertainty), it is the dominant systematic uncertainty for the $\zab=1.738$ absorber towards J1944$+$7705 because its spectrum was fully corrected for long-range distortions.

\subsubsection{Absorption profile modelling errors}\label{sss:moderr}

Like many previous MM analyses, we derived a single value of \daa\ from each absorber using the ``best'' model of the velocity structure, i.e.\ that with the lowest value of $\chi^2_\nu$. The best model was selected amongst several trial models which, by eye, did not appear to leave significant structures in the normalised residuals of individual transitions or the composite residual spectrum (CRS). This approach, coupled with the blinding of the spectrum (see \Sref{ss:genfit}), ensures an objective analysis. However, it does not guarantee that our selected model is the ``correct'' one, i.e.\ the most physically accurate model of the unblinded spectrum of that absorber. Therefore, we set aside several other trial models which yielded only marginally larger $\chi^2_\nu$ values and, after unblinding the spectrum, used these models to determine \daa\ in the same way as the main results presented in \Tref{t:res}. In all absorbers except the best-constrained one ($\zab=1.325$ towards J0120$+$2133), these alternative models yield \daa\ values that deviated from the main \daa\ result by substantially less than the statistical uncertainty ($\la$0.2$\sigma_{\rm stat}$). Therefore, in these cases, we regard any systematic modelling error as negligible.

However, for $\zab=1.325$ towards J0120$+$2133, one of our 3 alternative models yielded a value of \daa\ shifted by $-3.65$\,ppm relative to the main result, and with a substantially larger statistical uncertainty, $\sigma_{\rm stat}=5.50$\,ppm (cf.\ 3.45\,ppm). This model included 2 additional velocity components compared to the fiducial model; one either side of, and both within 2.5\,\kms\ of, the component fitted near a velocity of 0\,\kms\ in \Fref{f:J0120_1.3254}. The large increase in the statistical uncertainty likely indicates ``over-fitting'' -- that too many components have been used to fit this spectral feature, creating large degeneracies between parameters, including \daa. Another model, in which only a single additional component was fitted to the same spectral feature, was not viable: the additional component was rejected by {\sc vpfit} during the $\chi^2$ minimisation process, indicating that it was not statistically necessary in the fit. The other two viable alternative models gave \daa\ values that deviated from the main result by just 0.09 and $-0.58$\,ppm, both with similar $\sigma_{\rm stat}$ as the main result. \footnote{The first of these models added an \ion{Al}{iii} component corresponding to the weaker, broader of the two \ion{Mg}{ii} components at $\approx-55$\,\kms\ in \Fref{f:J0120_1.3254}; this component was rejected by {\sc vpfit} in other species. The second model added a further component in \ion{Mg}{ii}, \ion{Fe}{ii} and \ion{Al}{iii} $\approx$5\,\kms\ bluewards of the feature at $\approx-55$\,\kms; the additional component was rejected from \ion{Mg}{i}.} From these 3 viable, alternative models, we take the mean absolute deviation of \daa\ from the main result, 1.44\,ppm, as the systematic modelling uncertainty\footnote{If a large number of alternative models was found, all with very similar $\chi^2_\nu$ values, the RMS in \daa\ would be a more accurate choice for this uncertainty. However, with only 3 alternative models, we choose the mean absolute deviation from the main result as a more conservative estimator.}. This uncertainty is noted as an additional uncertainty for this absorber in \Tref{t:res}.

\subsubsection{Isotopic abundance variations}\label{sss:iso}

The fits to the absorption profiles that provided values of \daa\ in \Tref{t:res} included the isotopic and hyperfine components of the transitions, either from laboratory measurements or calculations \citep[reviewed in][]{Murphy:2014:388}, assuming that the absorption clouds have the same relative isotopic abundance patterns as the terrestrial environment \citep{Rosman:1998:1275}. However, if the absorbers have different isotopic abundances, the velocity centroids of the transitions we study will be slightly shifted relative to each other, potentially causing a systematic shift in our estimates of \daa\ derived from profile fits with terrestrial isotopic abundances. Although this is a well-known vulnerability of the MM method \citep[e.g.][]{Murphy:2001:1223,Murphy:2004:131}, there are no strong observational constraints on the isotopic abundances in quasar absorption clouds, so the effect on \daa\ is kept separate from the formal systematic uncertainty budget in previous works. We follow that approach here, but below we calculate the maximal effect on our measurements and consider this possible systematic effect when discussing our main results in \Sref{ss:mainres} and \Sref{s:disc} below, and in our conclusions (\Sref{s:conclusion}).

If the isotopic abundances are different in the absorbers, the largest effect on \daa\ is expected to arise from the Mg transitions used here: these have relatively large separations between the 3 isotopic components, $^{24,25,26}$Mg (up to $\approx$0.4\,\kms), both sub-dominant isotopes ($^{25,26}$Mg) have relatively large isotopic terrestrial abundance ($\approx$10\% each), and Mg transitions are very frequently observed and used in MM analysis at redshifts 0.3--1.8. For example, if the heavy Mg isotopes ($^{25,26}$Mg) were absent in an absorption system, fitting its \ion{Mg}{i/ii} lines with the terrestrial isotopic abundances transitions would yield a \daa\ that was too large by $\approx$5\,ppm on average \citep[when, as in typical MM analyses, the \ion{Mg}{i/ii} transitions are compared with the most common \ion{Fe}{ii} transitions;][]{Fenner:2005:468}. Our approach to assessing the impact of isotopic abundance variations on our results is to make a similar, extreme assumption for all atoms studied here. This is motivated by the chemical evolution model of star-forming galaxies of \citet{Fenner:2005:468} in which they found that, for all the elements studied here, the dominant isotope at terrestrial metallicities had even higher relative abundance in much lower metallicity environments. We expect that our quasar absorbers have sub-solar metallicities, like the vast majority of absorbers studied previously \citep[e.g.][]{Rafelski:2012:89,Jorgenson:2013:482,Cooper:2015:58,Lehner:2016:283,Glidden:2016:270}. Thus, for this test we simply assume that only the dominant terrestrial isotope of each element\footnote{Specifically, $^{24}$Mg, $^{28}$Si, $^{40}$Ca, $^{48}$Ti, $^{52}$Cr, $^{56}$Fe and $^{64}$Zn; Al and Mn have only a single stable isotope each.} is present in our absorbers and re-run our $\chi^2$ minimisation process starting from the \daa\ values in \Tref{t:res} to determine how \daa\ changes.

As expected, when using the dominant isotope only, we find negligible change in \daa\ for the two absorbers without \ion{Mg}{ii} fitted ($\zab=0.576$ towards J0120$+$2133 and $\zab=1.738$ towards J1944$+$2133), whereas the change is substantial for the other absorbers towards J0120$+$2133: $-5.3$, $-4.2$, $-7.8$ and $-1.9$\,ppm for $\zab=0.729$, 1.048, 1.325 and 1.343, respectively. The mean change in these 4 cases is $-4.8$\,ppm, consistent with the expectation from \citet{Fenner:2005:468} mentioned above. Unfortunately, removing the \ion{Mg}{ii} transitions from our analysis of these 4 absorbers leaves \daa\ only very weakly constrained -- the statistical uncertainties increase by factors of 2.7--4.5 from 3.5--18\,ppm to 8.3--84\,ppm -- which does not allow for a useful test of the effect of isotopic abundance variations. As with most previous MM constraints on \daa, we instead must state and interpret our results with this potential, astrophysical systematic error in mind.

\subsection{Statistical analysis of main results}\label{ss:mainres}

\begin{figure}
\begin{center}
\includegraphics[width=1.0\columnwidth]{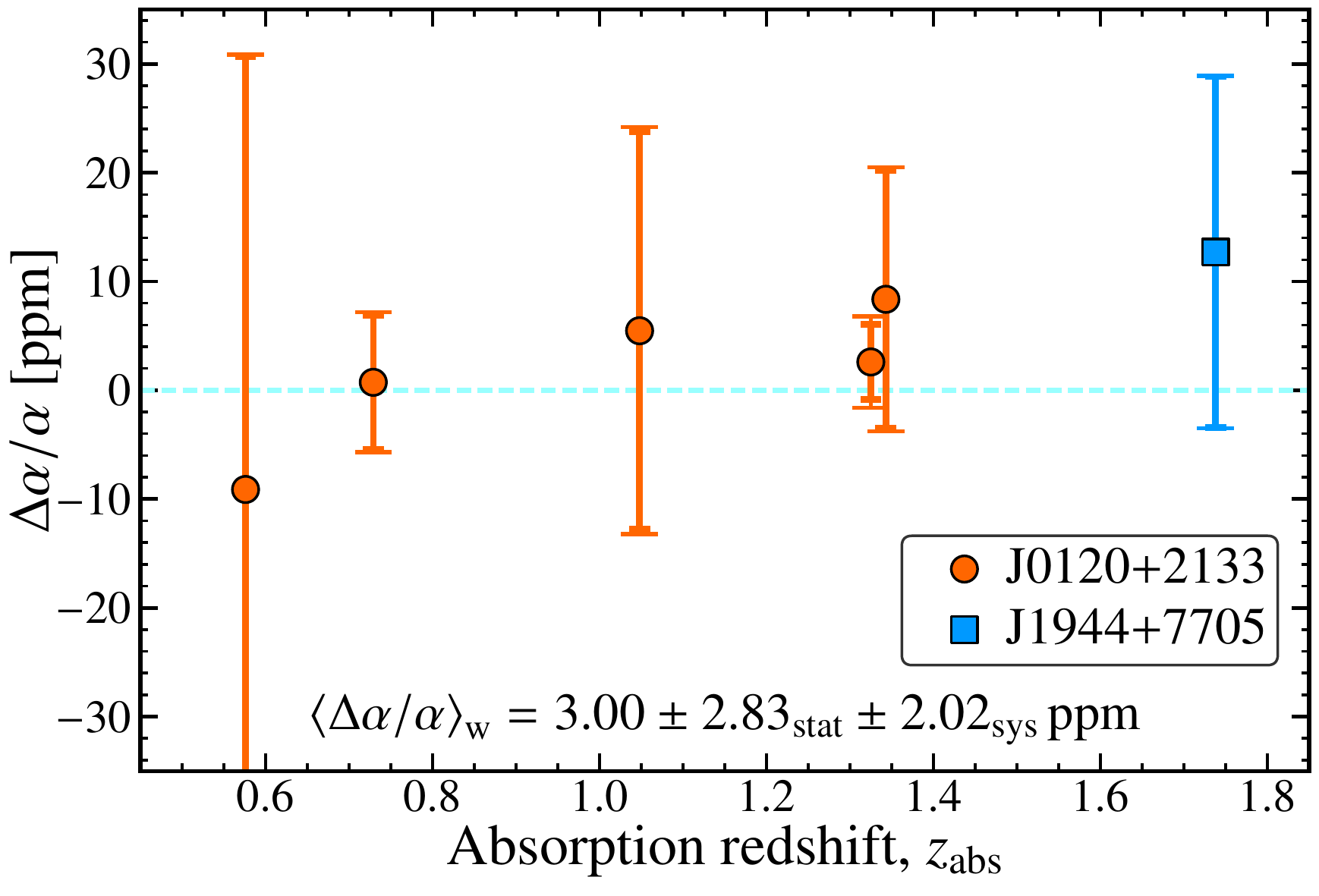}\vspace{-1em}
 \caption{\daa\ results with statistical and systematic uncertainties, for the 6 absorbers fitted in Figs.\ \ref{f:J0120_0.5764}--\ref{f:J1944_1.7384}, as a function of absorption redshift, \zab. The longer error bars with wider terminators represent the total uncertainties (quadrature sum of the statistical and systematic errors) whereas the 1$\sigma$ statistical errors are represented by the shorter error bars with narrower terminators (see \Tref{t:res}). The weighted mean \daa\ for the sample, $\left<\daa\right>_{\rm w}$ [\Eref{e:wmean}], taking into account the common-mode long-range distortion uncertainties for J0120$+$2133 (\Sref{sss:long}), is also shown.}
\label{f:da_vs_z}
\end{center}
\end{figure}

The best-fitting values of \daa, their 1$\sigma$ statistical uncertainties ($\sigma_{\rm stat}$) and the systematic error budget for each absorber are summarised in \Tref{t:res}. The total systematic uncertainty for each absorber, $\sigma_{\rm sys}$, is the quadrature sum of the individual systematic error components discussed in Sections \ref{sss:long}--\ref{sss:moderr}, and is smaller than the statistical uncertainty in all absorbers, as expected from the lack of scatter in the \daa\ values (see discussion in \Sref{ss:dares}). The long-range distortion uncertainties are dominant for the 5 absorbers towards J0120$+$2133 (due mainly to the 2012 exposures) and, for the absorber with the smallest $\sigma_{\rm stat}$ ($\zab=1.325$), the absorption profile modelling uncertainty of 1.44\,ppm is also important. \Fref{f:da_vs_z} shows all 6 results as a function of absorption redshift. Clearly, there is no evidence for a significant difference in $\alpha$ between the absorbers and the current laboratory value, or between individual absorbers, and no evidence ($<$0.6$\sigma$ significance) of a trend with redshift.

To form a weighted mean result from the 6 absorbers, we first find a weighted mean \daa\ for the J0120$+$2133 sight-line, recognising that the long-range distortion uncertainties cause correlated \daa\ uncertainties in its 5 absorbers, as discussed in \Sref{sss:long}. The weighted mean for J0120$+$2133, $\left<\daa\right>_{\rm w}=2.53\pm2.87_{\rm stat}$\,ppm, is calculated using the inverse sum of the variances from the statistical uncertainties and other systematic effects (i.e.\ intra-order distortions, redispersion and modelling uncertainties) as the weights. The change, $\sigma_{\rm LRD}$, in \daa\ caused by introducing an additional, postitive distortion slope to the J0120$+$2133 exposures (equal to the distortion uncertainty in each case) is shown in \Tref{t:res} (``LRD'' column)/ Note that is has both a magnitude and a sign (negative for one absorber but positive for all others). Its effect on the weighted mean was calculated as the difference of weighted means $\left|\left<\daa+\sigma_{\rm LRD}\right>_{\rm w}-\left<\daa\right>_{\rm w}\right|=1.73$\,ppm (``LRD'' column of the ``Ave.'' row in \Tref{t:res}). The contributions of the intra-order distortions to the J0120$+$2133 sight-line's systematic error budget is calculated as the quadrature difference $(\sigma^2_{\rm No\,LRD}-\sigma^2_{\rm No\,LRD\,or\,IOD})^{1/2}$, where $\sigma^2_{\rm No\,LRD}$ is the uncertainty in the weighted mean when using all error components except the long-rang distortion component in the weights, and $\sigma^2_{\rm No\,LRD\,or\,IOD}$ is similar but further excludes the intra-order distortion uncertainty component from the weights. The contributions from redispersion and modelling uncertainties are calculated in similar ways. These contributions are listed in their respective columns of the ``Ave.'' row in \Tref{t:res} (and, for the modelling uncertainty, its caption). They are summed in quadrature, together with the long-range uncertainty, to yield the total systematic uncertainty of $\sigma_{\rm sys}=2.10$\,ppm for the J0120$+$2133 sight-line.

The weighted mean \daa\ for the 6 absorbers is then simply that from the two independent sight-lines, i.e.\ the final two rows of \Tref{t:res}:
\begin{equation}\label{e:wmean}
\left<\daa\right>_{\rm w} = 3.00 \pm 2.83_{\rm stat} \pm 2.02_{\rm sys}\,{\rm ppm}\,.
\end{equation}
The systematic error component here was calculated from the quadrature difference of the uncertainties in the weighted means derived by including and excluding the sight-lines' systematic uncertainties in the weights. The main results, and the weighted mean above, all assume that the relative isotopic abundances in the absorption clouds are the same as the terrestrial ones. However, as discussed in \Sref{sss:iso}, we found that 4 of the 6 absorbers -- those in which \ion{Mg}{ii} contributes significantly to constraining \daa\ -- would have a significantly lower \daa\ ($\approx-4.8$\,ppm lower on average) if, instead, we assumed that only the dominant isotope of all elements was present. Under this extreme assumption, the weighted mean in \Eref{e:wmean} becomes
$\left<\daa\right>_{\rm w} = -3.35 \pm 2.83_{\rm stat} \pm 2.02_{\rm sys}$\,ppm, which is still consistent with no variation in $\alpha$.

Finally, we note that correcting the long-range distortions in the two quasar spectra made an important difference to the results above. For example, by not making these corrections to the spectra and determining \daa\ using the same fiducial velocity structures, we find that \daa\ differs by 0.4--2.0\,ppm from the main result for individual absorbers. The weighted mean result for J0120$+$2133 would be 1.8\,ppm lower, and that of J1944$+$7705 would be $1.3$\,ppm higher in this case; the overall weighted mean result have been lower by 1.7\,ppm if the distortion corrections had not been made. This is similar to some of the corrections found by \citet{Evans:2014:128} and \citet{Kotus:2017:3679}, though some of the measurements by the former had corrections of up to 10--15\,ppm. This illustrates the importance of correcting for the long-range distortions in individual exposures when attempting to measure \daa\ to part-per-million precision and accuracy.

\section{Discussion}\label{s:disc}

Our 6 new measurements of \daa\ contribute, with comparable precision, to the sample of 21 previous ``reliable'' measurements, i.e.\ those made using spectra corrected for long-range distortions \citep{Evans:2014:128,Kotus:2017:3679} or in systems that were resistant to this problem \citep{Murphy:2016:2461}. \citeauthor{Evans:2014:128} made 9 measurements in a single sight-line, from 3 different absorbers in 3 independent, distortion-corrected spectra (from Keck, Subaru and VLT), finding a weighted mean $\daa=-5.4\pm3.3_{\rm stat}\pm1.5_{\rm sys}$\,ppm, while \citeauthor{Kotus:2017:3679} made the highest-precision measurement from a single absorber, $\daa=-1.4\pm0.5_{\rm stat}\pm0.6_{\rm sys}$\,ppm, in a very high \SN\ distortion-corrected VLT spectrum. \citeauthor{Murphy:2016:2461} made 11 measurements using only the \ion{Zn}{ii} and \ion{Cr}{ii} transitions in 9 different absorbers (one per quasar in 6 Keck and 5 VLT spectra), finding a weighted mean of $\daa=0.4\pm1.4_{\rm stat}\pm0.9_{\rm sys}$\,ppm; the \ion{Zn/Cr}{ii} transitions have only a small wavelength separation, so these \daa\ measurements are resistant to long-range distortions. The weighted mean from our 6 Subaru absorbers in \Eref{e:wmean} is consistent with all these previous measurements within 1.7$\sigma$ (the previous measurements are consistent with each other within 1.5$\sigma$). The weighted mean from all 27 reliable measurements is
\begin{equation}\label{e:all}
\left<\daa\right>_{\rm w} = -1.0 \pm 0.5_{\rm stat} \pm 0.5_{\rm sys}\,{\rm ppm}\,,
\end{equation}
which is dominated by the high relative precision of the \citeauthor{Kotus:2017:3679} result. If that result is removed, the weighted mean of the remaining 26 measurements becomes $0.4\pm1.3_{\rm stat}\pm0.7_{\rm sys}$\,ppm. Neither of these results provide evidence for variations in $\alpha$.

Our new measurements also triple the sample and more than double the precision of \daa\ measurements from the Subaru telescope. The previous 3 measurements, along a single sight-line, by \citet{Evans:2014:128} had a weighted mean of $-11.2\pm7.8_{\rm stat}\pm2.4_{\rm sys}$\,ppm, which is consistent with our result in \Eref{e:wmean} at the 1.6$\sigma$ level. Combining these with our new measurements yields a weighted mean from all 9 Subaru measurements of $0.8\pm2.7_{\rm stat}\pm1.8_{\rm sys}$\,ppm. The weighted mean results from the distortion-corrected/resistant Keck and VLT measurements are $1.2\pm1.9_{\rm stat}\pm1.0_{\rm sys}$ (3 measurements from \citeauthor{Evans:2014:128}; 6 from \citeauthor{Murphy:2016:2461}) and $-1.5\pm0.5_{\rm stat}\pm0.6_{\rm sys}$\,ppm (3 measurements from \citeauthor{Evans:2014:128}; 5 from \citeauthor{Murphy:2016:2461}; 1 from \citeauthor{Kotus:2017:3679}), respectively. That is, the most reliable results from these 3 different telescopes are all consistent with no variation in $\alpha$ and also with each other within $<$2$\sigma$.

\begin{figure}
\begin{center}
\includegraphics[width=1.0\columnwidth]{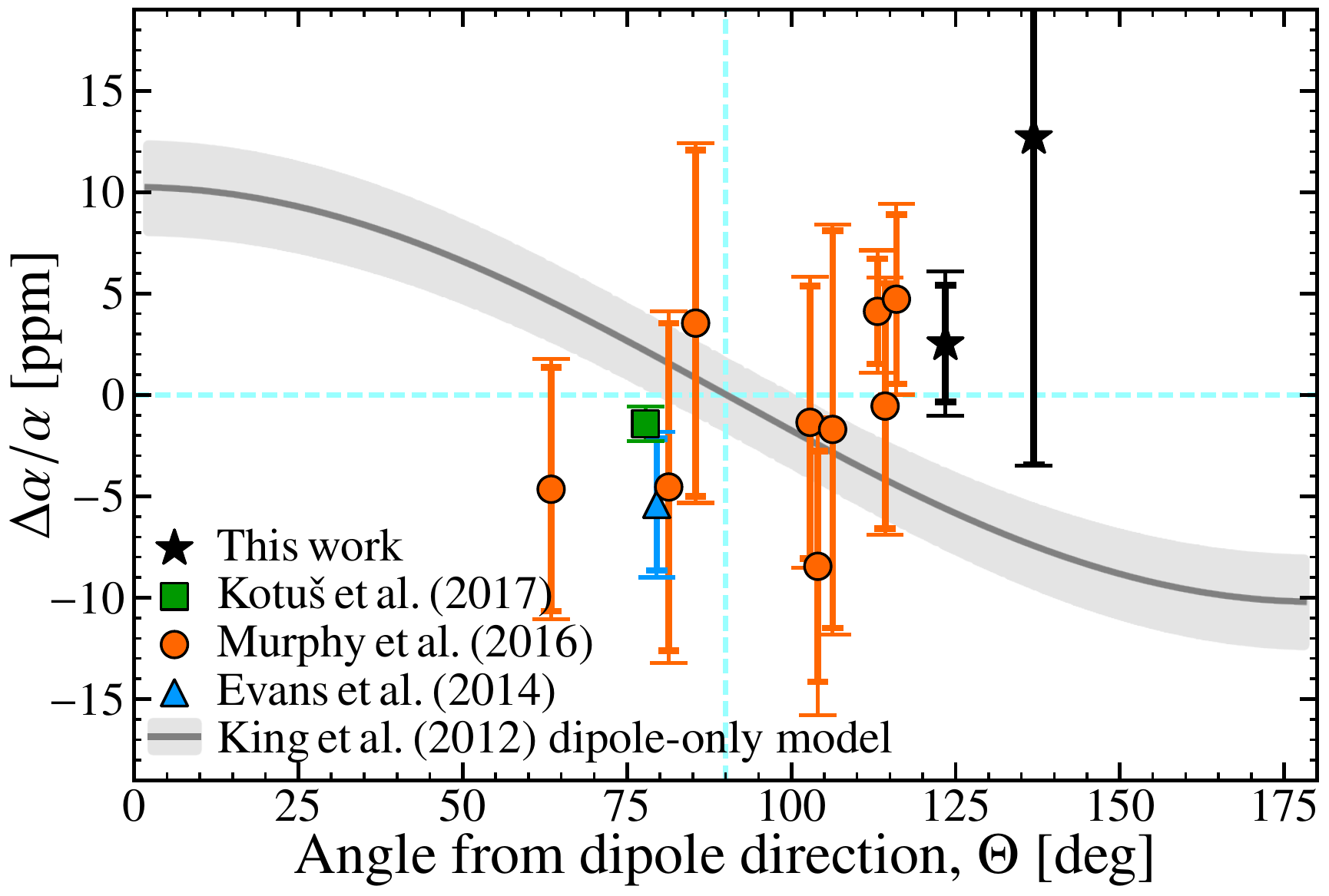}\vspace{-1em}
\caption{Recent measurements of \daa\ using optical spectra which were corrected for long-range distortions \citep[][and this work's results]{Evans:2014:128,Kotus:2017:3679}, or where \daa\ is insensitive to them \citep{Murphy:2016:2461}, as a function of angle on the sky away from the pole [at ${\rm (RA,Dec.)}=(17.4\pm0.9\,{\rm h},-58\pm9\,{\rm deg})$] of the \citet{King:2012:3370} dipole model, $\Theta$: $\daa=(10.2\pm2.1)\,\cos(\Theta)$\,ppm. The longer error bars with wider terminators represent the total uncertainties whereas the 1$\sigma$ statistical errors are represented by the shorter error bars with narrower terminators. The dipole expectation is shown as a solid line, with grey shading indicating the $\pm$1$\sigma$ uncertainty. Note that the 2 quasars studied here, J0120$+$2133 and J1944$+$7705, are the furthest from the dipole equator amongst this sample, enhancing their ability to test the dipole model. For J0120$+$2133 we plot the weighted mean \daa\ for the 5 absorbers studied here (see \Sref{ss:mainres} and \Tref{t:res}). A similar approach was taken for quasars in previous works with multiple measurements along their sight-lines (due to multiple absorbers and/or multiple spectra from different telescopes). A tabulation of all measurements in this figure is available in \citet{Murphy:2017:alphaSubaru}.}
\label{f:da_vs_theta}
\end{center}
\end{figure}

One of the motivations for targeting the two particular quasars studied here was their smaller separation from the anti-pole of the dipole-like variation in $\alpha$ across the sky implied by combining the large Keck and VLT samples of \daa\ measurements \citep{Webb:2011:191101,King:2012:3370}. \Fref{f:da_vs_theta} illustrates this point by comparing our \daa\ measurement for each sight-line (see \Sref{ss:mainres}) to the expected value from the ``dipole-only'' model of \citet{King:2012:3370} (see \Sref{s:intro}). Each data point in \Fref{f:da_vs_theta} is the weighted mean \daa\ for all absorbers, in all available spectra (from different telescopes), along a single quasar sight-line. It is apparent that, taken together, our 2 sight-line measurements do not align with the expected value of $\daa\approx-5$\,ppm -- they depart from it by $\ga$2$\sigma$ -- but that they are not highly inconsistent with it because of their total uncertainties of 3.6 and 16\,ppm (for J0120$+$2133 and J1944$+$7705, respectively)\footnote{Note that the dipole model was derived from \daa\ measurements which assumed terrestrial isotopic abundance ratios for all elements. It may be tempting to think that our result from J0120$+$2133 would be more consistent with the dipole expectation value if instead we assume that only the dominant isotopes were present in the absorbers, as discussed in \Sref{sss:iso}. However, the dipole model would also have to be modified for such a comparison to be valid.}. Nevertheless, when they are considered together with the other distortion-corrected/resistant measurements reviewed above, they certainly do not support the dipole model. The $\chi^2$ of all 13 measurements in \Fref{f:da_vs_theta} around the dipole model, using the statistical and systematic errors in the weights, is 43.5, which has a probability of just $\approx$0.003\% of being exceeded by chance alone, equivalent to rejecting the dipole model with 4.1$\sigma$ significance. However, to properly assess the consistency, or otherwise, of the dipole model with these data, the uncertainties in the model itself must also be taken into account. Also, even this simple significance estimate is again dominated by the high-precision, single-absorber measurement of \citet{Kotus:2017:3679}; removing this measurement decreases $\chi^2$ to 25.6 and increases the probability to 1.2\% (2.5$\sigma$). Therefore, we leave a more sophisticated analysis of this to a future paper, after a series of new measurements with distortion-corrected spectra become available (Kotu\v{s} et al., in prep.). Even if we cannot rule out the dipole model using the 27 reliable measurements made so far (and notwithstanding the likely susceptibility of the dipole evidence to long-range distortions, discussed in \Sref{s:intro}), their $\chi^2$ around $\daa=0$ is just 11.4, so $\daa=0$ is certainly preferred over the dipole model at present.

\section{Conclusions}\label{s:conclusion}

Given the susceptibility of all many-multiplet \daa\ measurements prior to 2014 to long-range wavelength distortions in the quasar spectra, obtaining a statistical sample of new measurements that are corrected for, or resistant to, this systematic error is an important goal. It is also imperative to build samples from all telescopes/spectrographs capable of high-precision \daa\ measurements to guard against, or discover, any currently unknown systematic effects that vary between them. Towards these goals, we obtained new, high-\SN\ ($\ga$50\,pix$^{-1}$ at 5000\AA) Subaru spectra of 2 quasars that, through supercalibration with ``attached'' solar twin observations, were corrected for long-range distortions (at least to first order). The spectral analysis was blinded (\Sref{ss:genfit}) and the best-fitting absorption profile model for each of the 6 absorbers studied (Figs.\ \ref{f:J0120_0.5764}--\ref{f:J1944_1.7384}) was selected with, and tested against, objective criteria. The spectra and fits to the absorption profiles, including all parameter estimates and uncertainties, are provided for full reproducibility in \citet{Murphy:2017:alphaSubaru}.

The analysis provided 6 new \daa\ measurements at $\zab=0.55$--1.75, which triples the available sample of Subaru results and improves their collective precision by more than a factor of 2. All 6 measurements are consistent with no change in $\alpha$ between the absorbers and the current laboratory value, there is no evidence of redshift evolution, and their weighted mean is also consistent with no variation: $\left<\daa\right>_{\rm w}=3.0\pm2.8_{\rm stat}\pm2.0_{\rm sys}$\,ppm (1$\sigma$ statistical and systematic uncertainties). The total precision of 3.5\,ppm is similar to that from 9 measurements by \citet[][3.6\,ppm]{Evans:2014:128}, but a factor of 2 poorer than the 11 \ion{Zn/Cr}-only measurements of \citet[][1.7\,ppm]{Murphy:2016:2461} and 4 times poorer than the single-absorber, highly precise constraint of \citet[][0.9\,ppm]{Kotus:2017:3679}.

The long-range distortions in our Subaru/HDS spectra have slopes similar in magnitude to those in previous works ($\approx 0$--2\,\msnm) and, had we not corrected for them, would have had an important impact on our \daa\ measurement, reducing it by 1.7\,ppm. Together with similar findings by \citet{Evans:2014:128} and \citet{Kotus:2017:3679}, this demonstrates the importance of correcting individual quasar exposures to ensure reliable \daa\ measurements. While statistical uncertainties dominated the error budget for all 6 absorbers, the largest systematic errors were from uncertainties in the long-range distortion corrections, uncertainties in the absorption profile model of our best-constrained absorber, and from redispersion of the quasar exposures onto a common wavelength grid (to enable combination into a single spectrum for analysis). Our results assume that the terrestrial isotopic abundances of the elements we study also prevail in the absorption clouds. This is a particularly important assumption for absorbers in which Mg transitions are used to constrain \daa, particularly the strong \ion{Mg}{ii} doublet. Making the extreme assumption that only the dominant terrestrial isotopes are present in the absorption clouds would lower \daa\ by 4.8\,ppm on average in the 4 absorbers with \ion{Mg}{ii}, and would lower the overall weighted mean to $-3.4$\,ppm (with the same total uncertainty of 3.5\,ppm).

Our 6 new measurements are consistent with the 21 previous distortion-corrected/resistant measurements \citep{Evans:2014:128,Murphy:2016:2461,Kotus:2017:3679}. The weighted means from the three different telescopes (Keck, Subaru and VLT) are also consistent with each other (and with $\daa=0$). Together, the 27 measurements have a weighted mean of $-1.0\pm0.5_{\rm stat}\pm0.5_{\rm sys}$\,ppm, indicating no variation in $\alpha$ at the 1\,ppm level. However, this statistic ignores the possibility that $\alpha$ varies across the sky, some evidence for which is implied if the previous large Keck and VLT samples are combined \citep[][noting that these measurements are likely to have been significantly affected by long-range distortions, \citealt{Whitmore:2015:446}]{Webb:2011:191101,King:2012:3370}. The two quasars studied here were selected because they lie closer to the anti-pole of the \citet{King:2012:3370} dipole model than the other 21 measurements, with the model expectation value of $\daa\approx-5$\,ppm. The weighted mean of our 6 new measurements departs from that expectation by $\ga$2$\sigma$. Considering the overall distribution of all 27 measurements (along 13 different sight-lines) on the sky, their departure from the dipole model is considerably more significant. A sophisticated estimate of the precise significance is required and is left to future work. Nevertheless, it is clear that our Subaru measurements, and the overall sample of 27 distortion-corrected/resistant values now available, do not support the dipole model.

\section*{Acknowledgements}

We thank Sr\dj{}an Kotu\v{s} for useful discussions about fitting the absorption profiles studied here. This work was based on data collected at Subaru Telescope, which is operated by the National Astronomical Observatory of Japan. The authors wish to recognise and acknowledge the very significant cultural role and reverence that the summit of Maunakea has always had within the indigenous Hawaiian community. We are most fortunate to have the opportunity to conduct observations from this mountain. MTM thanks the Australian Research Council for \textsl{Discovery Projects} grant DP110100866 which supported this work. KLC appreciates the observational support of R.\ Ponga, a University of Hawai`i (UH) at Hilo undergraduate student. The 2014 Subaru/HDS observing run was supported by a UH Hilo Research Council Seed Grant.















\bsp	
\label{lastpage}
\end{document}